\documentclass[12pt,preprint,showpacs,
preprintnumbers,nofootinbib]{revtex4}
\usepackage{epsf,graphicx}
\usepackage{amsmath}
\usepackage{times,pifont}

\def\beq{\begin{equation}}
\def\eeq{\end{equation}}
\def\bea{\begin{eqnarray}}
\def\eea{\end{eqnarray}}
\def\beqa{\begin{equation}\begin{array}{l}}
\def\eeqa{\end{array}\end{equation}}
\def\eqlab#1{\label{eq:#1}}
\def\figlab#1{\label{fig:#1}}

\def\barr{\left(\begin{array}{c}}
\def\earr{\end{array}\right)}
\def\bmat{\left(\begin{array}{cc}}
\def\emat{\end{array}\right)}

\def\Eqref#1{Eq.~(\ref{eq:#1})}

\def\Figref#1{Fig.~\ref{fig:#1}}

\def\sla#1{#1  \!\!\!\!\slash}

\def\slap{p \hspace{-2mm} \slash}

\def\half{\mbox{\small{$\frac{1}{2}$}}}

\def\thalf{\mbox{\small{$\frac{3}{2}$}}}

\def\third{\mbox{\small{$\frac{1}{3}$}}}

\def\al{\alpha}
\def\be{\beta}
\def\ga{\gamma} 
\def\de{\delta} \def\De{\Delta}
\def\veps{\varepsilon}  

\def\la{\lambda} \def\La{{\Lambda}}

\def\si{\sigma} \def\Si{{\it\Sigma}}
\def\th{\theta}  
\def\w{\omega}

\def\pa{\partial}

\def\pa{\partial}

\def\nn{\nonumber}

\def\Tau{{\mathcal T}}
\def\lag{{\mathcal L}}

\def\mathscr{\mathcal}

\def\3d{3-D}

\def\ol#1{\overline{#1}}

\def\ceft{$\chi$EFT}

\begin{document}
\preprint{ECT*-07-19}

\title{Chiral effective-field theory in the $\Delta$(1232) region :\\ 
II. radiative pion photoproduction}

\author{Vladimir Pascalutsa}
\email{vlad@ect.it}
\affiliation{European Centre for Theoretical Studies in Nuclear Physics 
and Related Areas (ECT*), Villa Tambosi,
 Villazzano, TN 38050,  Italy}

\author{Marc Vanderhaeghen}
\email{marcvdh@jlab.org}

\affiliation{Physics Department, The College of William \& Mary, Williamsburg, VA
23187, USA\\
Theory Center, Jefferson Lab, 12000 Jefferson Ave, Newport News, 
VA 23606, USA}

\date{\today}

\begin{abstract}
We present a theoretical study of
the radiative pion photoproduction on the nucleon
($\gamma N \rightarrow \pi N \gamma'$)
in the $\De$-resonance region, with the aim to determine the
magnetic dipole moment (MDM) of the $\Delta^+(1232)$.
The study is done within the framework of 
chiral effective-field theory where the expansion
is performed (to next-to-leading order) in the
$\delta$ power-counting scheme which is an extension of chiral
perturbation theory to the $\Delta$-resonance
energy region.
We present the results for
the absorptive part of the $\Delta$ MDM, as well as
perform a sensitivity study of the dependence of
$\gamma N \rightarrow \pi N \gamma'$ observables 
on the real part of the $\Delta$ MDM.
We find that an asymmetry for circular polarization of the photon beam
may provide a model-independent
way to measure the $\Delta$ MDM.
\end{abstract}


\maketitle
\thispagestyle{empty}

\section{Introduction}

The $\De(1232)$-isobar is the lowest nucleon excitation and
the most distinguished baryon resonance.
Unfortunately, its lifetime ($\approx 10^{-23}\, s$ )
is far too short for a direct measurement of its electromagnetic moments.
A measurement of the electromagnetic moments of such an unstable particle
can apparently be done only indirectly, in a
three-step process, where the particle is produced,
emits a low-energy photon, which plays the role
of an external magnetic field, and then decays.
In this way the magnetic dipole moment (MDM)
of $\Delta^{++}$ is accessed in 
the radiative pion-nucleon scattering ($\pi^+ p \to \pi^+ p \gamma$)~\cite{Nef78},
while the MDM of  $\Delta^+$ and $\De^0$ can be
accessed in the {\it radiative photoproduction}~($\gamma N \to \pi N \gamma^\prime$) of
neutral~\cite{Kotulla:2002cg}
and charged~\cite{charged} pions, on the proton and neutron respectively.

Some theoretical input is then needed in order
to extract the MDM value from observables. For instance,
the current Particle Data Group value of the $\De^+$ MDM~\cite{PDG2006}:
\beq
\mu_{\Delta^+} =  2.7 \mbox{${{+1.0} \atop {-1.3}}$}
(\mathrm{stat.}) \pm 1.5 (\mathrm{syst.}) \pm 3 (\mathrm{theor.})\, \mu_N\,,
\eeq
with $\mu _N=e/2M_N$ the nuclear magneton,
was obtained by the TAPS Collaboration at MAMI~\cite{Kotulla:2002cg}
using a phenomenological model
of the $\gamma p \to \pi^0 p \gamma^\prime$ 
reaction~\cite{Drechsel:2001qu}. The size of the error-bar is
rather large due to both experimental and theoretical uncertainties.

Recently a dedicated experimental effort
took place at MAMI using the Crystal Ball detector~\cite{Nstar07},
improving on the statistics of the TAPS data by almost
two orders of magnitude.
The present work is an effort to make an analogous improvement on the
theoretical side by using a more consistent and systematic framework of
chiral effective-field theory (\ceft). The first calculation
within \ceft \, of the
radiative pion-photoproduction  in the $\De$-resonance
region with the aim of extracting the $\De^+$ MDM has been presented
in a Letter~\cite{PV05}. In this more extended publication we present further
details and make a few improvements on the first work.

The previous studies of the radiative pion-photoproduction
were based on the so-called `effective Lagrangian approach'
\cite{Drechsel:2001qu,Machavariani:1999fr,Drechsel:2000um},
where one computes only the (unitarized) tree-level contributions,
and `dynamical models' \cite{Chiang:2004pw}, where some pion rescattering
effects are included in addition.

The framework of \ceft\ is more powerful in that it allows
to compute the chiral loop corrections in the consistent
framework of quantum field theory
(see, e.g.,~\cite{BKM,Ecker:1994gg,Scherer:2002tk,Kaplan:2005es} for reviews).
The chiral symmetry of the low-energy strong interaction, together
with other general principles, such as unitarity and
analyticity, is incorporated in \ceft\ to any given order
in a systematic expansion over the energy scales
and hadronic degrees of freedom.

In the problem at hand, the energy flow is defined
by the energies of incoming and outgoing photon.
To access the magnetic dipole moment of the $\De$
the energy of the incoming photon must be sufficient
to excite the resonance, while the emitted
photon must be soft. Therefore, in computing
this process, we use a chiral
expansion with $\De$-isobar degrees of freedom,
the so-called $\de$-expansion~\cite{PP03}, and simultaneously
the soft-photon expansion with respect to (w.r.t.)
the energy of the emitted photon. The soft-photon
expansion is performed to the next-next-to-leading order
(NNLO), since this is the order at which the MDM first appears.
The chiral expansion is performed to next-to-leading order (NLO).
In this way the present work can be viewed as an extension
of our recent NLO calculation of pion
photoproduction~\cite{Pascalutsa:2005ts,Pascalutsa:2005vq}
to the softly radiative pion photoproduction in the $\De$-resonance
region.

This paper is organized as follows.
In Sec.~\ref{sec2}, we define the electromagnetic moments of the
$\Delta(1232)$ and parameterize the $\gamma \Delta \Delta$ vertex. 
In Sec.~\ref{sec3}, we recall the relevant chiral Lagrangians
and explain the power counting in the $\de$-expansion
scheme, based on which
the leading- and next-to-leading-order contributions are then selected.
The chiral-loop contributions
to both the $\Delta$ self-energy and the $\gamma \Delta \Delta$-vertex
are evaluated in Sec.~\ref{sec4}.
In particular, we discuss how the $\Delta$ MDM acquires an imaginary
(absorptive) part due to the opening of the $\Delta \to \pi N$ decay channel.
After discussing the different observables for the
$\gamma N \to \pi N \gamma^\prime$ reaction in Sec.~\ref{sec5}, we
present results for these observables within the \ceft~framework in
Sec.~\ref{sec6}. We investigate in detail the sensitivity of different
observables on the $\Delta^+$ MDM, and discuss opportunities for
experiment. 
Section~\ref{sec7} summarizes the main points and conclusions of the paper.

\section{Electromagnetic moments of the $\Delta$}
\label{sec2}

Consider the coupling of a photon to a $\Delta$, \Figref{gadeldeltreevertex}. 
The matrix element of the electromagnetic current operator $J^\mu$
between spin 3/2 states can be decomposed into four multipole transitions:
a Coulomb monopole (C0), a magnetic dipole (M1),
an electric quadrupole (E2) and
a magnetic octupole (M3).
We first write its Lorentz-covariant decomposition which exhibits
manifest electromagnetic gauge-invariance:
\begin{eqnarray}
\langle \Delta(p') \,|\,e J^\mu(0) \,|\, \Delta (p) \rangle  &\,\equiv\,&\, 
\bar u_{\alpha}(p') \, \Gamma^{\alpha \beta \mu}_{\gamma \Delta \Delta} (p',p)
\, u_\beta(p) \nn\\
&\,=\,& -e \, \bar u_\alpha (p')\, \left\{ e_\De\,
F_1^\ast(Q^2) \, g^{\alpha \beta} \, \gamma^\mu \right.\nn \\
& + & \,\frac{i}{2M_\De}\left[ F_2^\ast(Q^2) \, g^{\alpha \beta}
+ F_4^\ast(Q^2) \,\frac{q^\al q^\be}{(2M_\De)^2}\right] \si^{\mu\nu} q_\nu \\
&+& \,\left. \frac{F_3^\ast(Q^2)}{(2M_\De)^2}\left[ q^\al q^\be \ga^\mu -
\half q\cdot\ga\,(g^{\al\mu}q^\be + g^{\be\mu}q^\al) \right] \, \right\}
 u_\beta(p) \, , \nn
\label{eq:gadeldeltree}
\end{eqnarray}
where $F^\ast_i$ are the $\gamma^* \Delta \Delta$ form factors,
$Q^2 \equiv - q^2 \geq 0$ is the photon virtuality with $q \equiv p' - p$,
$e_\Delta$ is the electric charge in units of $e$ (e.g., $e_{\Delta^+} = +1$),
such that $F^\ast_1(0)=1$,
$u_\al$ represents the spin-3/2 $\De$ vector-spinor.
The relation to the multipole
decomposition~\cite{Weber:1978dh,Noz90} can be written
in terms of the magnetic dipole ($\mu_\De$),
electric quadrupole ($Q_\De$) and magnetic octupole ($O_\De$) moments,
given by:
\begin{subequations}
\bea
\eqlab{EMmoments}
\mu_\De &=& \frac{e}{2 M_\Delta} \left[e_\De+F_2^\ast (0)\right]  \,,\\
Q_\De & = &  \frac{ e}{ M_\Delta^2}\left[ e_\De - \half F_3^\ast(0)\right]\,,\\
O_\De & = &  \frac{e}{2M_\Delta^3}
\left[ e_\De + F_2^\ast (0) - \half \left(F_3^\ast(0)+F_4^\ast(0)\right)\right]\,.
\eea
\end{subequations}

\section{Effective Lagrangian and power counting}
\label{sec3}

The construction of the SU(2) chiral Lagrangian with the
pion, nucleon, and $\De$-isobar degrees of freedom has
recently been reviewed~\cite{Pascalutsa:2006up,Bernard}.
Here we only specify the terms relevant to the
 NLO calculation of the $\ga N \to \pi N \ga^\prime$ process.
The Lagrangian is organized in orders of the
pion momentum and mass:
$\lag = \sum_i \lag^{(i)}$,
 with $i$ indicating the order.
In terms of the photon ($A_\mu$), pion ($\pi^a$),
nucleon ($N$), and $\De$-isobar ($\De_\mu$) fields we have:
\bea
\eqlab{lagran}
\lag^{(1)} & =& \ol N\,( i \sla{D} -{M}_{N}
+  \frac{g_A}{2f_\pi}\tau^a\sla{D}\, \pi^a 
\ga_5 )\, N
 + \ol\De_\mu \left(i\ga^{\mu\nu\al}D_\al -
M_\De\,\ga^{\mu\nu} \right) \De_\nu \,\nn\\
&+& \frac{h_A}{2 f_\pi M_\De} \left[ i
\ol N\, T^a \,\ga^{\mu\nu\la}\, (D_\mu \De_\nu)\, D_\la \pi^a
+ \mbox{H.c.}\right]\,, \nn\\
\lag^{(2)} & =&  \half D_\mu \pi^a \,D^\mu \pi^{a} - \half m_\pi^2 
\pi^2
+ \frac{3 e g_M}{2M_N (M_N + M_\Delta)}\,\left[i \, \ol N\, T^3
\,\pa_{\mu}\De_\nu \, \tilde F^{\mu\nu}  + \mbox{H.c.}\right]\\
&+& \frac{i e}{2M_\De}\, \ol \De_\mu
\,\half [ 1+\kappa^{(S)}_\De 
+ 3(1+\kappa^{(V)}_\De)\,\Tau_3] \,\De_\nu\, F^{\mu\nu}\,,\nn\\
\lag^{(3)} & =&
 \frac{-3 e\,g_E}{2M_N (M_N + M_\Delta)} \ol N \, T^3
\ga_5  (\pa_{\mu}\De_\nu)
 F^{\mu\nu}+ \mbox{H.c.},\nn
\eea
where $F^{\mu\nu}$ and $\tilde F^{\mu\nu}$
are the electromagnetic field strength and its dual;
$D_\mu$ is the e.m.\ covariant derivative defined for
the various fields as follows:
\bea
D_\mu \pi^a & = &\pa_\mu \pi^a + e  \,\veps^{ab3}A_\mu\pi^b\nn\\
D_\mu N &=&  \pa_\mu N - ie \half (1+\tau^3) A_\mu N \\
D_\mu \De_\nu &=& \pa_\mu \De_\nu -  ie \half(1+3\Tau^3) A_\mu\De_\nu \nn
\eea
The antisymmetrized
products of $\ga$-matrices are given by
$\ga^{\mu\nu}=\half[\ga^\mu,\ga^\nu]$,
$\ga^{\mu\nu\al}= i\veps^{\mu\nu\al\be}\ga_\be\ga_5$, etc.
The explicit form of the isospin matrices $T^a$ and  $\Tau^a$ has
been specified in our previous paper~\cite{Pascalutsa:2005vq}.

There are 8 dimensionless LECs  appearing in \Eqref{lagran}:
$g_A$, $h_A$, $g_M$, $g_E$, $\kappa_N^{(S)}$,
$\kappa_N^{(V)}$, $\kappa_\De^{(S)}$, and
$\kappa_\De^{(V)}$. To the considered order we may neglect
the difference between their values in the chiral limit
and at the physical pion mass. The physical
values of all, but the last two constants, are fairly
well known: $g_A\simeq 1.267$ is the axial
coupling of the nucleon; $h_A\simeq 2.85$ is the $\pi N\De$
coupling determined from the $\pi N$ scattering
$P$33 phase-shift~\cite{Pascalutsa:2006up};
$g_M$ and $g_E$ characterize the electromagnetic
$M1$ and $E2$ $N\to \De$ transition and are fixed from our fit of
the pion photoproduction
observables~\cite{Pascalutsa:2005vq}: $g_M = 2.97$, $g_E=-1$;
the isoscalar and isovector anomalous magnetic moments
of the nucleon are very well established:
$\kappa_N^{(S)}=0.12$,
$\kappa_N^{(V)}=3.7$.

The determination of $\kappa_\De^{(S)}$ and
$\kappa_\De^{(V)}$ --- the anomalous magnetic moments of the
$\De$ --- is precisely the aim of this work. These low-energy
constants will appear as the only free parameters in our calculations,
which will then be confronted with either the experimental or
lattice QCD data.

The thus introduced anomalous magnetic moments are defined as the deviation from the gyromagnetic ratio ($g=\mu/s$) from 2, the natural value for
an elementary particle of any spin $s$~\cite{Weinberg:1970bu,Ferrara:1992yc,Holstein:2006wi}.
In this notation the magnetic moment of the $\De$ corresponds
with\footnote{
It is understood that the magnetic moment of a resonance is a complex quantity.
(see the discussion of the imaginary part of the $\De$ MDM
in the following Section). Nonetheless, where not explicitly
specified we denote by $\mu_\De$ the real part of the $\De$ MDM.}
\bea
\mu_\De = \frac{e}{2M_\De} \left( 3 e_\De + \half \kappa^{(S)}_\De 
+ \thalf \kappa^{(V)}_\De \, \Tau_3 \,\right)\, ,
\eea
where $e_\De = (1+3\Tau_3)/2$ is the charge of the $\De$ in units of $e$.
Note that this definition of the anomalous magnetic moment of
the $\De$ has been recently introduced in \cite{Pascalutsa:2006up}
and is not widely used.
Usually one defines it as a deviation of the magnetic
moment from the magneton value: $e\,e_\De/(2 M_\De)$. Namely, in
\Eqref{EMmoments}, $F_2^\ast(0)$ corresponds precisely
with the conventional definition of the anomalous magnetic moment.
The relation between the two conventions is obvious:
$\kappa_\De^{(S,V)} = F_2^{\ast (S,V)}(0) -2\,$.
\newline
\indent
A first guideline as to the value of $\mu_\Delta$ is given by the
$1/N_c$ (with $N_c$ the number of colors) expansion  
of QCD proposed by 't Hooft~\cite{'tHooft:1973jz}
and Witten~\cite{Witten:1979kh}.
In the large-$N_c$ limit,
the baryon sector has an exact contracted
$SU(2 N_f)$ symmetry, where $N_f$ is the number of light quark flavors.
For three light quark flavors, this results in the
approximate SU(6) spin-flavor symmetry of QCD, which
underlies many quark model results. In this limit the nucleon and
$\Delta$ are degenerate, and their magnetic moments (for the same charge state)
are equal, i.e. $\mu_{\Delta^+} = \mu_p$.
Note that the $SU(6)$ relation
$\mu_{\Delta^+} = \mu_p \simeq 2.793$ (in nuclear
magnetons) yields very
different results for the $g$-value~:
$g_p = 5.58$, more than twice the value for an elementary particle,
whereas $g_{\Delta^+} = 2.44$, close to the natural value for an elementary
particle.

\subsection{Expansion about the resonance}
The chiral power counting in the presence of $\De$-isobar
field is also thoroughly discussed in the literature
(see Refs.~\cite{Pascalutsa:2006up,Bernard} for reviews).
Here we adopt the `$\de$ counting'
scheme~\cite{PP03}, where the $\De$-resonance excitation
energy:
\beq
{\it\De} \equiv M_\De -M_N \simeq 0.3 \,\mbox{GeV},
\eeq
is treated differently from the pion mass. One thus
has an expansion with two distinct light scales:
$m_\pi$ and  $\De$.
For power-counting purposes, one assumes
the following relation between the two:
\beq
\eqlab{pow}
\frac{m_\pi}{\La} \sim
\left(\frac{{\it\De}}{\La} \right)^2\,,
\eeq
where $\La \sim 1$ GeV stands for the scale
of chiral symmetry breaking, $4\pi f_\pi$, as well as
other heavy scales, such as the nucleon mass.
The power 2 in \Eqref{pow} is chosen  because it is the
closest integer power for the relation
between these parameters in nature.

The power counting is then done in orders of
$\de = {\it\De}/\La$, and the power counting
index of a given graph depends on whether the
generic momentum $p$ is in the {\it low-energy region}:
$p\sim m_\pi \sim \de^2$, or in the {\it resonance
region}: $p\sim {\it\De}\sim\de$. Such an energy-dependent
power counting turns out to be important for a
correct size estimate of the {\it one-Delta-reducible}
(ODR) graphs, which contain $\De$-isobar propagators
of the type:
\beq
S_{ODR} = \frac{1}{p-{\it\De}}\,.
\eeq
In the vicinity of the resonance these propagators diverge,
unless a whole class of ODR graphs is resummed, leading
to a ``dressed'' ODR propagator:
\beq
S_{ODR}^\ast = \frac{1}{p-{\it \De}-\Si}\,,
\eeq
where $\Si$ is the self-energy. The expansion for the
$\De$-isobar self-energy begins with $p^3$ and hence is
of order $\de^3$ in the resonance region. Therefore,
for
\beq
\eqlab{defofresreg}
|p-{\it\De} | \leq \de^3\,,
\eeq
the ODR propagators count as $\de^{-3}$.
The condition \Eqref{defofresreg} can be considered
as a stricter definition of the resonance region.

\subsection{Low-energy expansion}

The radiative pion photoproduction
experiments designed to measure the $\De$-resonance
magnetic dipole moment are done near the resonance energy
with the observation of a low-energy photon
in the final state.
This means the energy of the initial photon $\w$ is
in the resonance region, $\w \sim \de$, while the
energy of the final photon, $\w'$, is soft.
The expansion in $\w'$ can thus be done akin the
usual low-energy expansion (LEX), {\it i.e.}, the full amplitude
is expanded as:
\beq
\eqlab{lex}
f(\w,\w') = \frac{f_{-1}(\w)}{\w'} + f_0(\w) + f_1(\w)\,\w'
+\ldots\, = \sum_{i=-1} f_i \,{\w'}^i.
\eeq
The dependence of $f_i$'s on $\w$, $m_\pi$ and $\De$
can then be computed according to the $\de$-counting:
\beq
f_i = \sum_n f_i^{(n)}\, \de^n\,,\,\,\,\, \mbox{with}\,\,
\de \simeq \left\{ \w/\La,\,{\it\De}/\La,
\,\left(m_\pi/\La\right)^{1/2}\right\}\,.
\eeq

It is not difficult to see that the $\De$'s MDM
 starts to enter at order $\w'$. Therefore,
we shall need to have a complete calculation of
all three terms written out in \Eqref{lex} in order
to perform a model-independent extraction of
the MDM.

As far as the expansion in $\de$ is concerned,
we shall restrict ourselves to NLO.
We therefore begin with the pion photoproduction
at NLO, given by the graphs in \Figref{photopi}, where the blob
on the $\De$ propagator and the $\ga N\De$ vertex
indicate that those are corrected by the chiral loops as
shown in \Figref{loops}.
For the Born terms, only the electric coupling of the photon with energy
$\omega$ contributes to NLO.
Note also that to this order, only the imaginary
part of the loop corrections contributes, while the effect of the real parts
is limited to the renormalization of the apropriate
masses and coupling constants. This is why the analogous graphs
with the $\pi \De$ loops are omitted --- they do not produce
imaginary contributions in the $\De$-resonance region.

The minimal
insertion of the final photon in the NLO photoproduction amplitude
yields the radiative pion photoproduction
graphs in \Figref{diagrams2}, where the blob on the $\ga \De\De$ vertex
indicates the inclusion of the chiral corrections in \Figref{loops}(d) and (e).
The graphs in  \Figref{diagrams2} provide a complete NLO description of the
$f_{i}$ amplitudes in \Eqref{lex}, provided the
$\De$ MDM contribution from $\lag^{(2)}$ is included in the $\ga \De\De$
vertex.

To summarize, the graphs in \Figref{diagrams2}, with chiral corrections
\Figref{loops}, represent a complete calculation
of the radiative pion photoproduction amplitude to NNLO in $\w'$
and NLO in the $\de$-expansion.

\section{Chiral loops}
\label{sec4}

The chiral-loop corrections considered in this work
are comprised of the $\De$ self-energy \Figref{loops}(a),
$\ga N\De$-vertex \Figref{loops}(b,c) and $\ga\De\De$-vertex
\Figref{loops}(d,e) corrections.
To present the result of our calculations we introduce the
following dimensionless quantities:
\begin{subequations}
\bea
\al_p & = & (p^2+M_N^2-m_\pi^2)/2p^2, \\
\be_p & = & (p^2-M_N^2+m_\pi^2)/2p^2
\eea
\end{subequations}
which have the interpretation of an energy fraction carried by,
respectively, the nucleon and the pion in the loop (hence, $\al+\be =1$).
Furthermore, the corresponding 3-momentum fraction is given by
\bea
\la_p & = & \frac{1}{2 p^2}
\sqrt{p^2-(M_N+m_\pi)^2}\sqrt{p^2-(M_N-m_\pi)^2}\nn\\
&=&\sqrt{\al^2_p-\frac{M_N^2}{p^2}}
= \sqrt{\be^2_p-\frac{m_\pi^2}{p^2}}\,.
\eea

\subsection{Self-energy}
The result for the
$\De$ self-energy can be written as
\beq
\Si^{\al\be}_\De (p) = \frac{\slap}{p^2} \ga^{\al\be\si} p_\si\,
\Si_\De (\slap)\,,
\eeq
where $\Si_\De (\slap)$ has the Lorentz-form of a spin-1/2 self-energy,
in this case is given by
\beq
\Si_\De(\slap) =  \left( \frac{h_A}{8\pi f_\pi  }\right)^2
\frac{p^4}{M_\De^2}
 \!\int\limits_{-\al_p}^{1-\al_p}
\!dx \, (\al_p\,\slap+M_N)\, (x^2-\la^2_p) \,
\left[ l_p +
\ln(x^2-\la^2_p-i\veps)\right],
\eeq
where $l_p = -2/(4-d) + \ga_E -1 - \ln (4\pi{\it \La}^2/p^2) $,
with $d\simeq 4$ the number of dimensions,
$\ga_E \simeq 0.5772 $ the Euler constant,
and ${\it \La}$ the renormalization scale in dimensional
regularization.

Note that on the mass-shell one simply has,
\beq
\ol u_\al(p) \,\Si^{\al\be}_\De (p)\,u_\be(p) = -
\ol u_\al(p) \,\Si_\De (\slap)\,u^\al(p) = \Si(M_\De)
\eeq
where $u_\al$ is the vector-spinor of the free $\De$, solution
of the following equations: $(\slap - M_\De)u_\al(p) = 0 = p^\al u_\al(p)
= \ga^\al u_\al(p)$, $\ol u_\al \,u^\al = -1$.
 
We use the on-mass-shell renormalization, such that
$M_\De$ is the physical mass and the field
renormalization constant satisfies: Re$\, Z_\De = 1$.
This amounts to introducing a counterterm
which subtracts the first two terms
in the expansion of the self-energy around the
physical mass,
\beq
\eqlab{expansion}
\Si_\De (\slap) = \Si_\De(M_\De)+ \Si_\De'(M_\De)
\, (\slap - M_\De)+\ldots\,
\eeq
We subtract the real parts only. Subtracting
the imaginary part of the self-energy would violate unitarity.

Working at NLO in the EFT expansion, we neglect
completely the self-energy contributions beyond
the orders written explicitly in~\Eqref{expansion}.
Thus, the only contributions that survive the
renormalization and expansion are the following ones:
\begin{subequations}
\eqlab{sigmas}
\bea
\mathrm{Im}\,\Si_\De(M_\De) & = &
\frac{\pi}{2} M_\De C^2 \!\int\limits_{-\la}^{\la}
\!dx \, (\al+r)\, (x^2-\la^2) \nn\\
&=& - (2\pi/3) M_\De C^2 (\al+r)\, \la^3, \\
\mathrm{Im}\,\Si_\De' (M_\De) &= & \frac{\pi}{2}\, C^2 \!\int\limits_{-\la}^{\la}
\!dx \,(x+\al)\, [x^2-\la^2-2(\be-x) (x+\al+r)] \nn\\
& = & - 2\pi\, C^2\,
\la \left[ \al\be  (\al+r)
- \third \la^2 (r+r^2-\mu^2)\right],
\eea
\end{subequations}
where $C=M_\De h_A/(8\pi f_\pi)$,
$\al = \al(M_\De^2)$, $\be = \be(M_\De^2)$,  $\la = \la(M_\De^2)$.
The first term in the expansion
has the usual intepretation in terms of the width:
Im$\,\Si_\De = -\Gamma_\Delta /2$.
Using the empirical value of $\Gamma^{(exp)}_\Delta \simeq 115$ MeV, we deduce
the value of $h_A \simeq 2.85$.

The resulting $\Delta$ propagator can then be written as~:
\begin{eqnarray}
S(p)^{\mu \nu} = \frac{1}{Z_\Delta}
\frac{1}{\left(\gamma \cdot p - \bar M_\Delta \right)}
\left\{ - g^{\mu \nu} + \frac{1}{3} \gamma^\mu \gamma^\nu
+ \frac{1}{3 \bar M_\Delta} \left( \gamma^\mu p^\nu - \gamma^\nu p^\mu \right)
+ \frac{2}{3 \bar M_\Delta^2} p^\mu p^\nu \right\},
\end{eqnarray}
where the wave function renormalization constant $Z_\Delta$
and complex mass $\bar M_\Delta$
are given by~:
\begin{eqnarray}
Z_\Delta &=& 1 - i \, {\rm Im} \Sigma^\prime_\Delta (M_\Delta),  \\
\bar M_\Delta &=& M_\Delta - i \, \Gamma_\Delta / 2Z_\Delta.  
\end{eqnarray}

\subsection{The absorptive MDM}

Considering the chiral-loop corrections to the $\ga \De\De$-vertex
[see \Figref{loops}(d), (e)]  we shall focus on the imaginary parts only.
As mentioned above, the real parts can be renormalized away to the considered
order. Certainly this is possible only when we work in the vicinity of
the resonance and for the physical pion masses. The behavior of the
$\De$ MDM away from the physical pion mass, will be discussed in a
forthcoming publication.

It is convenient to write the general Lorentz-structure of the
chiral correction to the electromagnetic current of the $\De$
in the following form:
\bea
\left<\De'|J^\mu|\De\right>&= &
-e\, \bar u_\al (p')  \left[  \ga^\mu\, F(q^2)  + \frac{(p'+p)^\mu}{2M_\De}
\, G(q^2) \right]  u^\al (p) \,,
\eea
where $q=p'-p$ is the photon momentum, $u^\al$
is the Rarita-Schwinger vector-spinor, $F$ and $G$ are electromagnetic form factors.
We have omitted the electric quadrupole and magnetic octupole terms
here because
they enter at higher order in the LEX, \Eqref{lex}.
In this notation, the correction to the MDM is given by $e F(0)/2M_\De$.
In what follows we limit ourselves to the case of real photons, $q^2=0$.

The contribution of the graphs \Figref{loops}(d) and (e)
to the absorptive part of the electromagnetic form factors
can be written, respectively, as~\cite{PV05}:
\begin{subequations}
\bea
\mathrm{Im}\, F^{(d)} &=& -\pi C^2 \!\int\limits_{-\la}^{\la}
\!dx \,(x+\al) \,[ \,\half (x^2-\la^2)-(x+\al) (x+\al+r)\,]\,\nn\\
& = & 2\pi C^2 \,\la \,\left[\al^3 + r (\al^2+\la^2) +2\la^2\al\right]\,,\\
\mathrm{Im}\, G^{(d)} &=& \pi C^2 \!\int\limits_{-\la}^{\la}
\!dx \,(x+\al) (\mu^2-r^2-2x)(x+\al+r)\,,\\
 \mathrm{Im}\, F^{(e)} & = & -\pi C^2 \!\int\limits_{-\la}^{\la}
\!dx \,(x+\be) \,[ \,x^2-\la^2-\half (1+\be + x+r)^2+2(x+\be)^2 \,]\,\nn\\
&=& -(4\pi/3) C^2 \la \left[-2(1-2r)\la^2 + 3\al(1+\la^2)
+3\al^2(\al-2) +3r\mu^2 \right]\,,\\
\mathrm{Im}\, G^{(e)} & =& -2\pi C^2 \!\int\limits_{-\la}^{\la}
\!dx \,(x+\be)^2 \,(\al-x+r)\,,
\eea
\end{subequations}
with $C$, $\al$, $\be$ and $\la$ defined below \Eqref{sigmas}.

The electromagnetic gauge symmetry imposes
an intricate relation among a linear combination of
these functions and the self-energy. Namely, writing
the Ward-Takahashi
identity for each of the charge states of the $\De$ we
obtain:
\bea
\eqlab{WTIrels}
&& \De^{++}:  \, F^{(d)}+G^{(d)}+F^{(e)}+G^{(e)}
= 2(1-\Si_\De')\,,\nn\\
&& \De^{+}:  \, \third (F^{(d)}+G^{(d)})+
\mbox{$\frac{2}{3}$}(F^{(e)}+G^{(e)})
= 1-\,\Si_\De'\,,\nn\\
&& \De^{0}:  \,-\third (F^{(d)}+G^{(d)})+\third (F^{(e)}+G^{(e)})
= 0 \,,\\
&& \De^{-}: -\,(F^{(d)}+G^{(d)})
= -(1-\Si_\De')\,,\nn
\eea
where 
\beq
\Si_\De'=\pa/\pa \slap \, \Si_\De(\slap)|_{\slap=M_\De}
\eeq
is the first derivative of the $\De$ self-energy.
In deriving these relations we have assumed that the vertex correction
and the self-energy are computed to the same order
and are renormalized in the same scheme.
Since for now we are interested only in the absorptive parts
we shall not be addressing here the details of renormalization.

Using the empirical values for the couplings and masses
(e.g., $C\simeq 1.51$), we find
\bea
\mathrm{Im}\, F^{(d)} &\simeq & 2.63\,,\nn\\
\mathrm{Im}\, G^{(d)} &\simeq & -1.98\,,\nn\\
 \mathrm{Im}\, F^{(e)} & \simeq &  1.11\,,\\
\mathrm{Im}\, G^{(e)} & \simeq & -0.45\,,\nn \\   
\mathrm{Im}\, \Si_\De' &\simeq & -0.655\,\nn,
\eea
and hence can verify the relations \Eqref{WTIrels} numerically.
The absorptive part of the $\De$-resonance MDM is then found to be
[in units of $(e/2M_\De)$]:
\bea
\eqlab{predict}
&& \mathrm{Im}\, \mu_{\De^{++}}  =  \mathrm{Im}\,
[ F^{(d)} + F^{(e)} ]  \simeq 3.74 \,,\nn\\
&& \mathrm{Im}\, \mu_{\De^{+}}  =  \third \,\mathrm{Im}\,
[ F^{(d)} + 2 F^{(e)} ] \simeq 1.62 \,,\nn\\
&& \mathrm{Im}\, \mu_{\De^{0}}  =  \third\,\mathrm{Im}\,
[ - F^{(d)} + 2 F^{(e)} ]  \simeq -0.14 \,,\\
&& \mathrm{Im}\, \mu_{\De^{-}}  =  -\,\mathrm{Im}\, F^{(d)}
\simeq -2.63 \,.\nn
\eea
We checked our result in two independent ways~: firstly by calculating the
loop integrals as outlined above and taking their imaginary parts,
and secondly by directly calculating the corresponding
phase space integrals, using Cutkosky rules. We checked analytically that
both calculations give the same result for the imaginary part of the different
$\Delta$ MDMs.  

We note here that numerically our results are different from
an analogous EFT calculation of Hacker et~al.~\cite{Hacker:2006gu}.
Their result for the isoscalar and isovector
absorptive magnetic moments is
given by ~\cite{Hacker:2006gu}:
$\mathrm{Im}\, \mu^{(S)}_{\De} \equiv 
\mathrm{Im}\, ( \mu_{\De^{+}} + \mu_{\De^{-}}) =0$, and
$\mathrm{Im}\, \mu^{(V)}_{\De} \equiv 
\mathrm{Im}\, ( \mu_{\De^{+}} - \mu_{\De^{-}})
= 3 \times 0.37 \,(e/2M_\De) = 1.115 \, (e/2 M_\De)$.
In contrast we have:
$\mathrm{Im}\, \mu^{(S)}_{\De} \simeq -1$ and 
$\mathrm{Im}\, \mu^{(V)}_{\De} \simeq 4.25$, in units of $\De$ magnetons.
The discrepancy arises mainly because of
additional expansions in $m_\pi$ and ${\it \De}$
performed in~\cite{Hacker:2006gu}, which are justified from
the point of view of the power-counting used in that work.
Besides those expansions,
the numerical value of $\mathrm{Im}\, \mu^{(V)}_{\De}$ is further
reduced in~\cite{Hacker:2006gu}
due to a smaller value of the $\pi N \Delta$-coupling.
The value
$h_A^2 \simeq 8.12$ used in our calculations
(corresponding with $\Gamma_\Delta = 0.115$ GeV),
 must be compared with the
coupling $(2 g)^2 \simeq 5.08$ used in Ref.~\cite{Hacker:2006gu}.

We conclude this section with a
word on the interpretation of the absorptive MDMs. They apparently
quantify the change in the width
of the resonance that occurs in an external magnetic field $B$:
\beq
\De \Gamma = 2 \, \mathrm{Im}\, \mu_{\De} \, \vec B\cdot\vec n_s\,,
\eeq
where $\vec n_s$
is the direction of the resonance's spin, see Ref.~\cite{Binosi:2007ye}
for more details. Equivalently, one may look for a change in the
lifetime of the resonance:
\beq
\De \tau/\tau = -2 \, \mathrm{Im}\, \mu_{\De} \, \vec B\cdot\vec n_s\,\tau,
\eeq
where $\tau = 1/\Gamma$ is the lifetime.
Such a change in the lifetime appears to be extremely small
in moderate magnetic fields and is difficult to be observed directly.
On the other hand,
there is perhaps a possibility to compute the absorptive MDMs
of hadron resonances in lattice QCD
where the effect of arbitrarily large magnetic
fields on the width can in principle be studied.
 
Also, in the reaction such as the radiative pion photoproduction
there are sensitivities on the absorptive part of the
$\De$ MDM, which in the future could lead to its independent determination from
experiment. Until then, we are compelled to use the ChEFT predictions
\Eqref{predict} in our subsequent analysis of this reaction.

\section{Observables for the $\gamma N \to \pi N \gamma^\prime$ reaction}
\label{sec5}

To access the MDM of the $\Delta^+$, we consider the
radiative pion photoproduction process on a nucleon, i.e., the
$\gamma N \to \pi N \gamma^\prime$ reaction.
In this section we define a few useful observables of
this reaction.

In the discussion of observables it is customary to
denote the incoming photon energy by $E_\ga$ (instead of $\w$ as above)
and the outgoing photon energy by $E_\ga'$ (instead of $\w'$).
In the rest of the paper we shall adhere to this notation.
Furthermore,
all cross sections will be given in the center-of-mass
({\it c.m.}) system
of the initial $\gamma N$ state. All kinematical
quantities will also be quoted
in the {\it c.m.}\ system, except for the
incoming photon energy which we traditionally denote by its {\it lab}
system value $E_\gamma^{lab}$.
\newline
\indent
The $\gamma N \to \pi N \gamma^\prime$ reaction cross section
$d \sigma / d E_\gamma^\prime d \Omega_\pi d \Omega_\gamma$  
is five-fold differential w.r.t.\ to the outgoing photon energy
$E_\gamma^\prime$, and w.r.t.\ the solid angles of outgoing pion and photon.
Because measuring a five-fold differential cross section with sufficient
precision is very challenging, we shall also consider several partially
integrated cross sections. For example,
$d \sigma / d E_\gamma^\prime d \Omega_\pi$
denotes the $\gamma N \to \pi N \gamma^\prime$
cross section differential w.r.t.\ the outgoing photon energy and the pion
solid angle, where one has integrated over all outgoing photon directions. 
\newline
\indent
It has been shown in Ref.~\cite{Chiang:2004pw} that
in the soft-photon limit (i.e., for outgoing photon energies
$E_\gamma^\prime \to 0)$, the
$\gamma N \to \pi N \gamma^\prime$ observables exhibit a low energy
theorem. It is therefore useful to
introduce the ratio~\cite{Chiang:2004pw}:
\begin{eqnarray}
\label{eq:R1}
  R \,\equiv \, \frac{1}{\sigma_\pi} \cdot
  E^\prime_\gamma
  \frac{d\sigma}{dE^\prime_\gamma} ,
\end{eqnarray}
where $d\sigma / dE^\prime_\gamma$ is the
$\gamma N \to \pi N \gamma^\prime$
cross section integrated over the pion and photon angles, and
$\sigma_\pi$ is the angular integrated cross section
for the $\gamma N \to \pi N$ process weighted with the bremsstrahlung
factor, as detailed in Ref.~\cite{Chiang:2004pw}.
This ratio has the property that the
low-energy theorem demands:  $R \to 1$, as $E_\ga'\to 0$ (the soft-photon
limit).
\newline
\indent
Besides the unpolarized cross section, we shall also consider here the
linear photon polarization asymmetry $\Sigma^\pi$, defined as~:
\begin{eqnarray}
\Sigma^\pi = \frac{\left(d \sigma_\perp / d E_\gamma^\prime d \Omega_\pi \right) -
\left(d \sigma_\parallel / d E_\gamma^\prime d \Omega_\pi \right) }{\left(d
\sigma_\perp / d E_\gamma^\prime d \Omega_\pi \right) +
\left(d \sigma_\parallel / d E_\gamma^\prime d \Omega_\pi \right) },
\label{eq:linsigmapi}
\end{eqnarray}
where the superscript $\pi$ in
$\Sigma^\pi$ indicates that the plane ($\Phi_\pi = 0^\circ$)
is defined by the
incoming photon and outgoing pion directions, and 
where $d \sigma_\perp$ ($d \sigma_\parallel$) are 
the cross sections for perpendicular (parallel) initial photon polarizations
respectively w.r.t.\ the above-defined plane.
\newline
\indent
Analogously one can define a linear photon polarization asymmetry
$\Sigma^\gamma$ as~:
\begin{eqnarray}
\Sigma^\gamma = \frac{\left(d \sigma_\perp / d E_\gamma^\prime d \Omega_\gamma \right) -
\left(d \sigma_\parallel / d E_\gamma^\prime d \Omega_\gamma \right) }{\left(d
\sigma_\perp / d E_\gamma^\prime d \Omega_\gamma \right) +
\left(d \sigma_\parallel / d E_\gamma^\prime d \Omega_\gamma \right) },
\label{eq:linsigmaga}
\end{eqnarray}
where the superscript $\gamma$ in $\Sigma^\gamma$
indicates that the plane ($\Phi_\gamma = 0^\circ$)
is defined by the
incoming photon and outgoing photon directions, and 
where $d \sigma_\perp$ ($d \sigma_\parallel$) are 
the cross sections for perpendicular (parallel) initial photon polarizations
respectively w.r.t. the above defined plane.
\newline
\indent
For circularly polarized photons the photon asymmetry vanishes for a two body
reaction due to reflection symmetry with respect to the reaction plane.
However, for a three body final state the circular photon polarization 
asymmetry is finite
and can be obtained from the cross section for circular photon polarization.
When defining the plane spanned by the incoming photon and the outgoing pion
directions as reference plane (i.e., taking $\Phi_\pi = 0^\circ$),
the outgoing photon azimuthal angular
dependence can be expanded as~:
\begin{eqnarray}
d \sigma_{\lambda_\gamma} =
d \sigma_{0}
+ \sum_{n=1} \, \left[ \cos (n \Phi_\gamma) \, d \sigma_{n}+ \lambda_\gamma \,
\sin (n \Phi_\gamma) \, d \sigma_{n}' \right],
\label{eq:polcross}
\end{eqnarray}
where $d \sigma$ is a short-hand notation for
the {\it c.m.}\ cross section
$d \sigma / d E_\gamma^\prime d \Omega_\pi d \Omega_\gamma$, 
while $\lambda_\gamma = \pm 1$ denotes the photon helicity.
Denoting the cross sections for the two different photon helicities
by $d \sigma_\pm$, one obtains the unpolarized cross section as:
\begin{eqnarray}
\half \left( d \sigma_+ + d \sigma_- \right) =
d \sigma_{0}
+ \cos \Phi_\gamma \, d \sigma_{1}
+ \cos 2 \Phi_\gamma \, d \sigma_{2} + ...,
\label{eq:unpolexp}
\end{eqnarray}
and the difference in photon helicity cross sections as:
\begin{eqnarray}
d \sigma_+ - d \sigma_-  =
2 \left[ \sin \Phi_\gamma \, d \sigma_{1}'
+ \sin 2 \Phi_\gamma \, d \sigma_{2}' + ... \right]. 
\label{eq:polexp}
\end{eqnarray}
\indent
One can then define a circular photon polarization asymmetry
(integrated over the photon polar angle) as:
\begin{eqnarray}
\Sigma^\pi_{circ} (\th_\pi) \equiv \frac{2}{\pi} \,
\frac{\int_{-1}^{+1} d \cos \theta_\gamma \; \int_0^{2 \pi} d \Phi_\gamma \,
2 \, \sin \Phi_\gamma \;
\left(d \sigma_+   -
      d \sigma_-  \right) }{
\int_{-1}^{+1} d \cos \theta_\gamma \; \int_0^{2 \pi} d \Phi_\gamma \;
\left(d \sigma_+   +
      d \sigma_- \right) } = \frac{2}{\pi}
\frac{\int_{-1}^{+1} d \cos \theta_\gamma \,d \sigma_{1}'}{\int_{-1}^{+1} d \cos \theta_\gamma \,d \sigma_{0}} \,,
\label{eq:casymm}
\end{eqnarray}
where the superscript $\pi$ indicates that the
reference plane ($\Phi_\pi = 0^\circ$)
is defined by the incoming photon and outgoing pion
directions~\footnote{The prefactor $2/\pi$ in Eq.~(\ref{eq:casymm})
arises by defining the asymmetry originally as the ratio of $(d \sigma_+ - d
\sigma_-)/(d \sigma_+ + d \sigma_-)$ and integrating numerator and denominator
over the upper hemisphere only. This is equivalent with the definition of
Eq.~(\ref{eq:casymm}) which involves the $\sin \Phi_\gamma$ moment and where
numerator and denominator are integrated over the complete
photon solid angle.}.   
\newline
\indent
Similarly we define the circular photon polarization
asymmetry for fixed
outgoing-photon direction and integrated pion direction:
\newline
\indent
\begin{eqnarray}
\Sigma^\gamma_{circ} (\th_\ga)\equiv \frac{2}{\pi}
\frac{\int_{-1}^{+1} d \cos \theta_\pi \; \int_0^{ 2 \pi} d \Phi_\pi \,
2 \, \sin{\Phi_\pi} \;
\left(d \sigma_+   -
      d \sigma_- \right) }{
\int_{-1}^{+1} d \cos \theta_\pi \; \int_0^{2 \pi} d \Phi_\pi \;
\left(d \sigma_+   +
      d \sigma_-  \right) },
\label{eq:casymmg}
\end{eqnarray}
where the superscript $\gamma$ 
indicates that the reference plane ($\Phi_\gamma = 0^\circ$) is defined by the
incoming photon and outgoing photon directions,
while the pion solid angle is being fully integrated over.

\section{Results and discussion}
\label{sec6}

We now present the numerical results of our \ceft\
calculation for the
$\gamma N \to \pi N \gamma^\prime$ observables.
We begin with
\Figref{compeft} which shows the EFT results for
the ratio $R$ of the $\gamma p \to \pi^0 p \gamma^\prime$
differential cross section relative to the $\gamma p \to \pi^0 p$
one as defined in Eq.~(\ref{eq:R1}),
for the linear photon polarization asymmetry $\Sigma^\pi$ defined in
Eq.~(\ref{eq:linsigmapi}), and for the circular photon polarization asymmetry
$\Sigma^\pi_{circ}$ defined in Eq.~(\ref{eq:casymm}). The dashed
curves represent our previous results \cite{PV05}, while
the solid lines show the result of the present EFT calculation.

The difference
between the two calculations is that in Ref.~\cite{PV05} the outgoing
photon energy was counted as $O(\de^2)$, while presently
we use the soft-photon expansion, \Eqref{lex}, which required the inclusion
of additional bremsstrahlung diagrams.
The present calculation thus includes all graphs shown in \Figref{photopi} and
\Figref{loops}, while in the previous calculation
non-resonant terms (i.e., graphs without $\Delta$'s) were not
included. 
Note that while we now include the
 non-resonant graphs with the initial photon coupled to the charge of the
nucleon or pion, we do not include
the coupling of the initial photon to the anomalous magnetic moment of the
nucleon (in the non-resonant graphs), as it is
suppressed by one more power of $\de$ compared to the electric
coupling.

Both calculations in \Figref{compeft} are shown for $F_2^\ast(0) = 0$
corresponding with $ \mu_{\Delta^+} = 1$ (in $\Delta$ magnetons). 
One sees that the effect
of the non-resonant contributions included in the new calculation
mainly affect the linear and circular photon asymmetries.
For values of $E_\gamma^\prime$ above about 70 MeV,
the non-resonant terms in the
present EFT calculation  are responsible for the 
significant decrease in $\Sigma^\pi$ in contrast to the
previous calculation,
where the linear photon asymmetry is nearly constant
over the whole $E_\gamma^\prime$ range shown in \Figref{compeft}.  
The effect of the non-resonant terms also shows up in a sizable change in
the circular photon asymmetry $\Sigma^\pi_{circ}$ with increasing
outgoing photon energy.

In the soft-photon limit ($E_\gamma^\prime \to 0$), the present
calculation of $\gamma N \to \pi N \gamma^\prime$ goes to
 the corresponding
NLO photoproduction result~\cite{Pascalutsa:2005vq}.
The latter has been shown \cite{Pascalutsa:2006up} 
to be in an agreement with the well-measured $\gamma p \to \pi^0 p$
observables in a 100 MeV window 
around the $\Delta$-resonance position.
In particular, one sees the nice agreement with the
experimental $\gamma p \to \pi^0 p$ linear photon asymmetry, shown by the
data point for $\Sigma^\pi$ in \Figref{compeft}.
\newline
\indent
In \Figref{gapio_tot},
we show the outgoing photon energy dependence for the absolute
$\gamma p \to \pi^0 p \gamma^\prime$
differential cross sections for different values of $\mu_{\Delta^+}$. 
The cross sections in \Figref{gapio_tot}
have been integrated over all pion and outgoing photon
angles and are compared with the first data for this reaction from
Ref.~\cite{Kotulla:2002cg}. One sees that at $E_\gamma = 350$~MeV,
all calculations are consistent with the available data.
With increasing photon energies
(compare $E_\gamma = 400$~MeV vs.\ $450$~MeV in \Figref{gapio_tot}), 
the present EFT calculation yields an increasing reduction in the
absolute cross section relative to the previous  calculation.
Although this trend is consistent with the existing data,
both the precision and sensitivity of these integrated data
are insufficient to extract a value of $\mu_{\Delta^+}$ besides
favoring values in the range between 1 and  3 $\Delta$
magnetons. This is also demonstrated in \Figref{eftang}, where the outgoing
photon angular dependence of the cross section is shown when integrated over
all pion angles and for all values of $E_\gamma^\prime \geq 30$~MeV.
The EFT calculation is consistent with the angular dependence displayed
by the first data of Ref.~\cite{Kotulla:2002cg}, and the higher energy
results ($E_\gamma = 450$~MeV) also seem to favor
the range $\mu_{\Delta^+} \simeq 1 - 3$.
\newline
\indent
To enhance the sensitivity to the value of $\mu_{\Delta^+}$, we next study
$\gamma p \to \pi^0 p \gamma^\prime$
observables that are more differential. In \Figref{eftpi} we show the 
cross section and linear photon asymmetry $\Sigma^\pi$, both
differential in the outgoing photon energy and pion angle for
three values of the outgoing pion angle. One verifies that in the low-energy
limit, $E_\gamma^\prime \to 0$, the photon asymmetries of the present
EFT calculation are in very good
agreement with the photon asymmetry data for the
$\gamma p \to \pi^0 p$ reaction as required by the low-energy theorem.    
With increasing photon energies, the linear photon asymmetries in the
present EFT calculation steadily decrease and show an increasing sensitivity
to the value of $\mu_{\Delta^+}$, especially around
$\theta_\pi^{c.m.} = 90^o$.

The $\gamma p \to \pi^0 p \gamma^\prime$ 
differential cross sections show a pion 
angular behavior which can mainly be understood from 
the following two observations. 
First, the underlying
$\gamma p \to \pi^0 p$ pion angular distribution is peaked at 
$\theta_\pi^{c.m.} = 90^o$ being approximately symmetrical around this value
in the resonance region. Second,
the forward-backward asymmetry arises from the
bremsstrahlung processes which display a strong enhancement for backward
pion angles compared with forward pion angles, as noted in
Ref.~\cite{Chiang:2004pw}. One furthermore notices from \Figref{eftpi} that,
for photon energies $E_\gamma^\prime \geq
70$~MeV, the unpolarized
differential cross sections also show a modest sensitivity
to the value of $ \mu_{\Delta^+}$.
\newline
\indent
For an increased sensitivity to the $\Delta^+$ MDM, it was suggested in
Ref.~\cite{Chiang:2004pw} to measure the asymmetry for a
circularly polarized photon beam, as defined
in Eqs.~(\ref{eq:casymm}-\ref{eq:casymmg}). These helicity
asymmetries are only non zero
for a three-body final state and vanish in the soft-photon limit
$E_\gamma^\prime \to 0$, i.e, when the $\gamma N \to \pi N \gamma^\prime$
process reduces to the two-body, $\gamma N \to \pi N$ process.

In \Figref{phieft}, we show the NLO EFT results for the outgoing photon
azimuthal-angle dependence of the $\gamma p \to \pi^0 p \gamma^\prime$
cross sections, for $\Phi_\pi = 0^\circ$ (i.e., the initial photon and
outgoing pions define the reference plane).
We show both unpolarized cross section
(upper panels) and the difference in helicity cross sections for a circularly
polarized photon beam. 

The azimuthal-angle dependencies of 
the helicity-averaged and helicity-difference 
cross sections follow the expansions of
Eqs.~(\ref{eq:unpolexp}) and (\ref{eq:polexp}) respectively.
One sees from \Figref{phieft} that the helicity difference cross section is
dominated by its lowest harmonic, i.e. the term proportional to
$\sin \Phi_\gamma$ in Eq.~(\ref{eq:polexp}). Hence, a measurement of the
$\Phi_\gamma$ dependence of the helicity difference cross section
allows to extract its lowest moment, defined as the circular 
photon polarization asymmetry $\Sigma^\pi_{circ}$, through
Eq.~(\ref{eq:casymm}). 
\newline
\indent
The EFT predictions for the
circular photon polarization asymmetries
$\Sigma^\pi_{circ}$  and
$\Sigma^\gamma_{circ}$ are shown in
\Figref{helpi} and \Figref{helga} respectively.
The NLO EFT calculation firstly verifies that
$\Sigma^\pi_{circ}$ and $\Sigma^\gamma_{circ}$
vanish in the limit $E_\gamma^\prime \to 0$.
With increasing photon energy, these asymmetries show a very clear sensitivity
on the $\Delta^+$ MDM. One sees from \Figref{helpi} that, for
$E_\gamma = 400$~MeV and around $E_\gamma^\prime = 80$~MeV,
the asymmetry changes by 0.1 to 0.2 when varying the $\Delta^+$ MDM
in the range $\mathrm{Re} \, \mu_{\Delta^+} = 1 - 5$ of $\Delta$ magnetons.
Note also that for a given value of $\mathrm{Re} \, \mu_{\Delta^+}$
both $\Sigma^\pi_{circ}$ and $\Sigma^\gamma_{circ}$ change sign between
forward and backward angles.
The sensitivity of the circular photon polarization asymmetries on
$\mathrm{Re} \, \mu_{\Delta^+}$ can be understood from the
observation that these observables arise from the
product of the imaginary part of the (resonant)
$\gamma p \to \pi^0 p \gamma^\prime$ amplitude with the
real part of the $\gamma p \to \pi^0 p \gamma^\prime$ amplitude.
Hence the sensitivity on the real part of the $\Delta^+$ MDM is amplified
by multiplication with the large imaginary part, mainly arising from the
bremsstrahlung diagrams containing one (resonant)
$\Delta$ line, where the final
photon is radiated from either initial or final proton lines.   
The photon helicity asymmetry $\Sigma^\pi_{circ}$ is also sensitive to the
imaginary part of the $\Delta^+$ MDM by multiplication with the real part
of the $\gamma p \to \pi^0 p \gamma^\prime$ amplitude, as is shown by the
dotted curves in \Figref{helpi} and \Figref{helga}, which have been obtained
by switching off the imaginary part of $F_2^\ast(0)$.
Note that in the present EFT calculation the imaginary part of the $\Delta$
MDM is obtained as a
prediction of the one-loop calculations of \Figref{loops} (d) and (e), as
detailed in Sec.~\ref{sec4}.
\newline
\indent
We explore the large sensitivity of $\Sigma^\pi_{circ}$ on
$\mathrm{Re}\, \mu_{\Delta^+}$ further in \Figref{scirc}. The top panel of
\Figref{scirc} illustrates that for a photon beam energy of
$E_\gamma = 350$~MeV, where the initial $\Delta$ is near resonant,
there is a range of about 30-40 MeV in final photon energy where the
photon helicity difference cross section $\sigma_+ - \sigma_-$
is near linear in $E_\gamma^\prime$. The bottom panel of
\Figref{scirc} which is obtained from the top panel by dividing through
$E_\gamma^\prime$ shows this slope in $E_\gamma^\prime$. 
One sees that the slope is directly proportional to
$\mathrm{Re}\, \mu_{\Delta^+}$. Although this comes out of the NLO EFT calculation,
we like to point out that this is a model independent feature, resulting
from the low-energy theorem for the $\gamma N \to \pi N \gamma^\prime$
process. Indeed
in the low-energy expansion of \Eqref{lex} the dependence on the
$\Delta^+$ MDM comes in at the linear term
in $E_\gamma^\prime$. Because $\Sigma^\pi_{circ}$ has to vanish (exactly)
in the limit $E_\gamma^\prime \to 0$, 
it starts out with the linear term in the LEX, which
also depends linearly on the $\Delta^+$ MDM.
\newline
\indent
In view of the large sensitivity of $\Sigma^\pi_{circ}$ on the
$\Delta^+$ MDM, it will
be of interest to compare with first data for this observable.  
Such data have recently been taken by the Crystal Ball @ MAMI
$\gamma p \to \pi^0 p \gamma^\prime$ experiment~\cite{Nstar07}. Although this
experiment was optimized for a linearly polarized photon beam, the
experiment also took data with a circularly polarized photon beam.
\Figref{scirc} illustrates that a future dedicated measurement of this photon
helicity asymmetry for $E_\gamma^\prime \leq 40$~MeV provides
an opportunity for a model independent extraction of the $\Delta^+$ MDM.
\newline
\indent
Finally we also show the NLO EFT results for the
$\gamma p \to \pi^+ n \gamma^\prime$ reaction in
\Figref{eftpip} and \Figref{helpip}. 
By comparing \Figref{eftpi} with \Figref{eftpip}, one sees that the
ratio of the $\gamma p \to \pi^+ n \gamma^\prime$ to
$\gamma p \to \pi^0 p \gamma^\prime$ reactions is
around a factor 6, around
$\theta_\pi^{c.m.} = 90^\circ$. The origin of the enhancement in the
charged pion channel is the bremsstrahlung from the light charged pion.
However the sensitivity to the resonance properties
in the charged pion observables seems to be lesser pronounced
than in the neutral pion channel.
Nonetheless, the charged pion observables may provide further
test of the present EFT framework in which both channels
are calculated simultaneously with the same set of parameters.

\section{Conclusion}
\label{sec7}

In this work we continue to develop and apply a systematic extension
of chiral perturbation theory to the $\De$-resonance region.
Here we have presented the calculation of the imaginary part
of the $\De$ magnetic dipole moment (MDM), as well as a
comprehensive study of the observables
of the $\gamma N \to \pi N \gamma^\prime$ process 
in the $\Delta(1232)$-resonance region.

The chiral expansion for the $\gamma N \to \pi N \gamma^\prime$
amplitudes is performed
using the so-called $\delta$ power-counting scheme,
to next-to-leading order (NLO).
Furthermore, we use a simultaneous low-energy expansion in the
energy $\omega^\prime$ of the emitted photon which
includes all terms up to O($\omega^\prime$), where the
$\De$ MDM first appears.
To this order, besides the tree-level resonant diagrams, the
calculation involves one-loop vertex corrections to the $\gamma N \Delta$ and
$\gamma \Delta \Delta$ vertices, as well as non-resonant (Born) diagrams.
The only free parameter which enters at this order is the
$\Delta$ MDM. The one-loop corrections to the
$\gamma \Delta \Delta$ vertex give an imaginary (absorptive) part to the
$\Delta$ MDM, which was quantified.

Concerning the observables, the outgoing-photon energy
and angular dependencies
of the $\gamma p \to \pi^0 p \gamma^\prime$ unpolarized cross section were
found to be consistent with first experimental data for this process
(assuming about 10 percent uncertainty of our results, coming from the
neglect of higher-order contributions).

We quantified the sensitivities of the
$\gamma p \to \pi^0 p \gamma^\prime$ cross section and photon asymmetries 
to the $\Delta^+$ MDM. It appears that, at low energies of the outgoing
photon, the dependence of the cross-section and linear-photon
asymmetries on the MDM is quadratic, i.e., depends on $|\mu_\De|^2$. 
The asymmetry for a circularly
polarized photon beam, however, displays a linear dependence on the
$\Delta^+$ MDM. The helicity difference $\gamma N \to \pi N
\gamma^\prime$ cross section for a circularly polarized photon beam
vanishes linearly with $\omega^\prime$ when approaching the soft-photon limit,
its slope being proportional to the $\Delta^+$ MDM. Therefore, a
dedicated measurement of the $\gamma p \to \pi^0 p \gamma^\prime$
cross sections with a circularly polarized photon beam provides a unique
opportunity for a model-independent extraction of the $\Delta^+$ MDM. 


\begin{acknowledgments}

This work is supported in part by DOE grant
DE-FG02-04ER41302 and contract DE-AC05-06OR23177 under
which Jefferson Science Associates operates the Jefferson Laboratory.  
The work of V.~P.\ is partially supported  by the European Community-Research Infrastructure Activity under the FP6 "Structuring the European Research Area" programme (HadronPhysics, contract RII3-CT-2004-506078).

\end{acknowledgments}

\newpage

\begin{figure}[t,b,p]
\centerline{
\epsfxsize=5cm
\epsffile{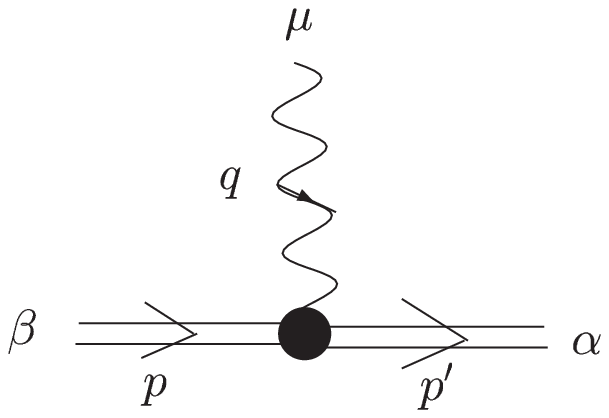}
}
\caption{The $\gamma \Delta \Delta$ vertex. The four-momenta of 
the initial (final) $\Delta$ and of the photon are given by 
$p$ ($p^\prime$) and $q$ respectively. 
The four-vector indices of the initial (final) 
spin 3/2 fields are given by $\beta$ ($\alpha$), and 
$\mu$ is the four-vector index of the photon field.}
\figlab{gadeldeltreevertex}
\end{figure}

\begin{figure}
\includegraphics[width=0.9\columnwidth]{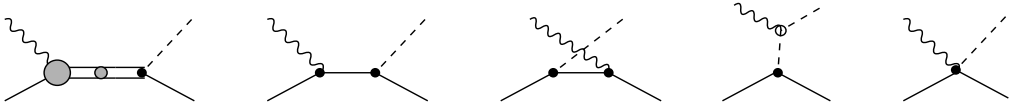} 
\caption{Diagrams for the $\gamma N \to \pi N $ reaction 
at NLO in the $\delta$-expansion, considered in this work. 
Double lines represent the $\De$ propagators.}
\figlab{photopi}
\end{figure}

\begin{figure}
\includegraphics[width=0.9\columnwidth]{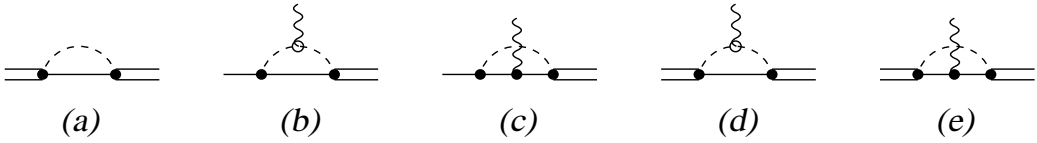} 
\caption{Chiral loop corrections considered in this work. 
Double lines represent the $\De$ propagators.}
\figlab{loops}
\end{figure}

\begin{figure}
\includegraphics[width=0.8\columnwidth]{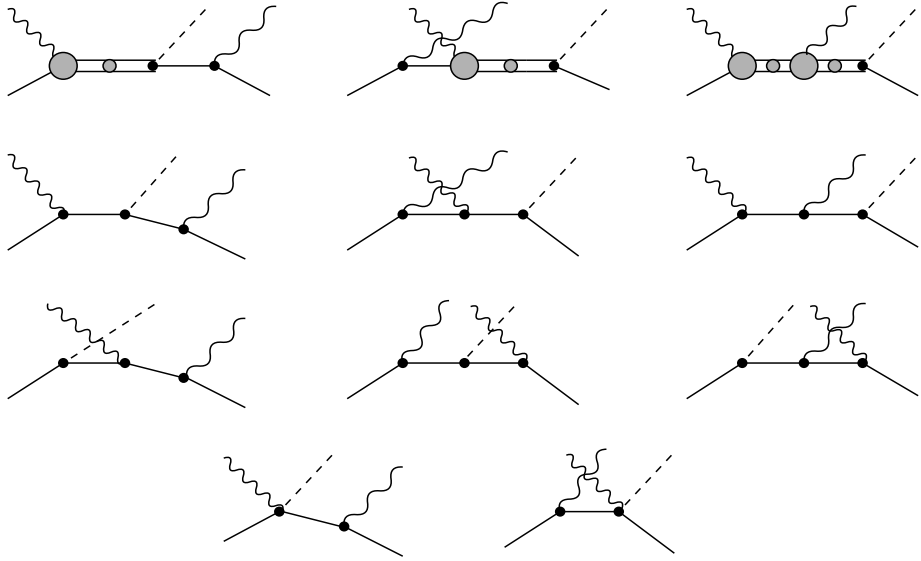} 
\caption{Diagrams for the $\gamma p \to \pi^0 p \gamma^\prime$ reaction 
at NLO in the $\delta$-expansion, 
considered in this work. Double lines represent the $\De$ propagators. 
For the $\gamma p \to \pi^+ n \gamma^\prime$ reaction, 
the pion-pole diagrams where either of the two photons couples to the charged 
pion are taken into account.}
\figlab{diagrams2}
\end{figure}

\begin{figure}[t,h]
\includegraphics[width=0.7\columnwidth]{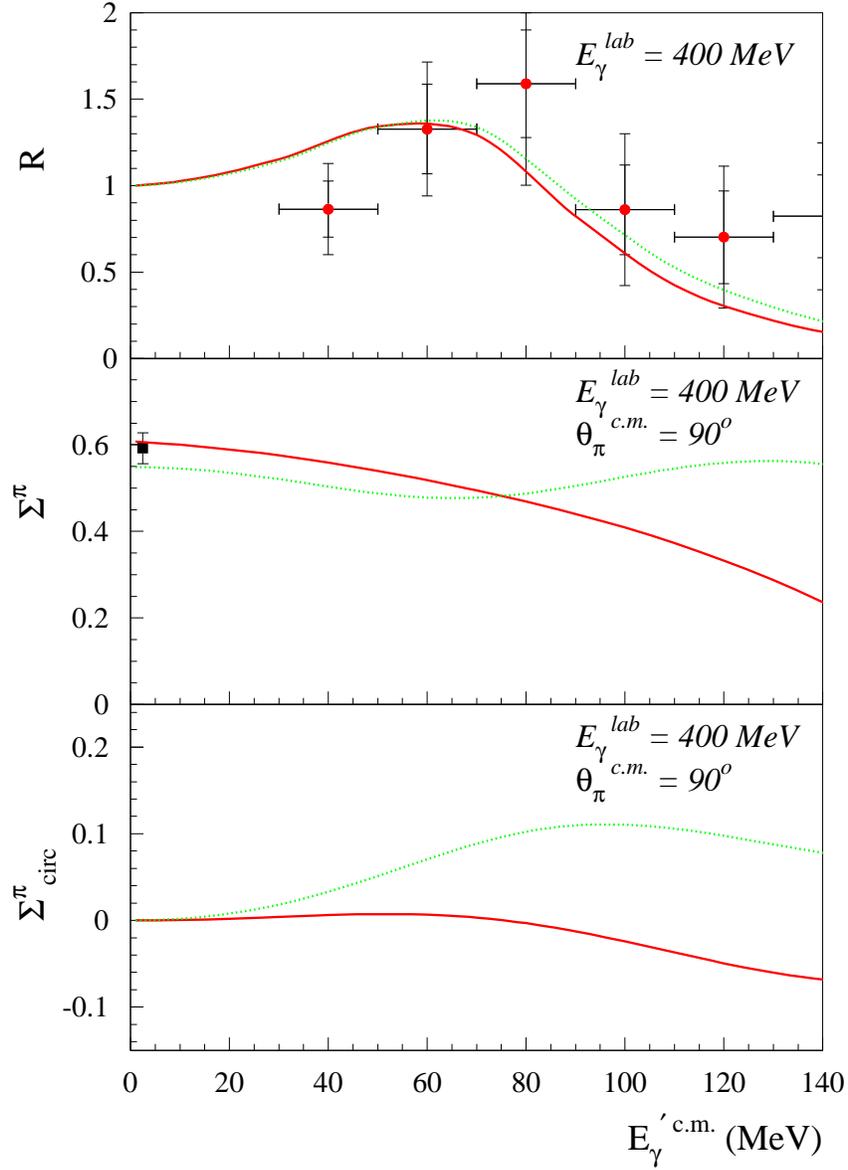}
\caption{
(Color online) Outgoing photon energy dependence 
for different observables (cross section ratio $R$ of Eq.~(\ref{eq:R1}), 
linear photon asymmetry $\Sigma^\pi$, and  
circular photon asymmetry $\Sigma^\pi_{circ}$) for the 
resonant EFT calculation 
of Ref.~\cite{PV05} (green dotted curves) in comparison  
with the present EFT results (red solid curves),  
both for $\mathrm{Re} \, \mu_{\Delta^+} = 1$. 
The data are from Ref.~\cite{Kotulla:2002cg}, and the linear photon 
asymmetry data point for $E_\gamma^\prime = 0$ is
$\gamma p \to \pi^0 p$ photon asymmetry from 
Refs.~\cite{Beck:1999ge,Leukel01}. 
}
\figlab{compeft}
\end{figure}

\begin{figure}[t,h]
\includegraphics[width=0.7\columnwidth]{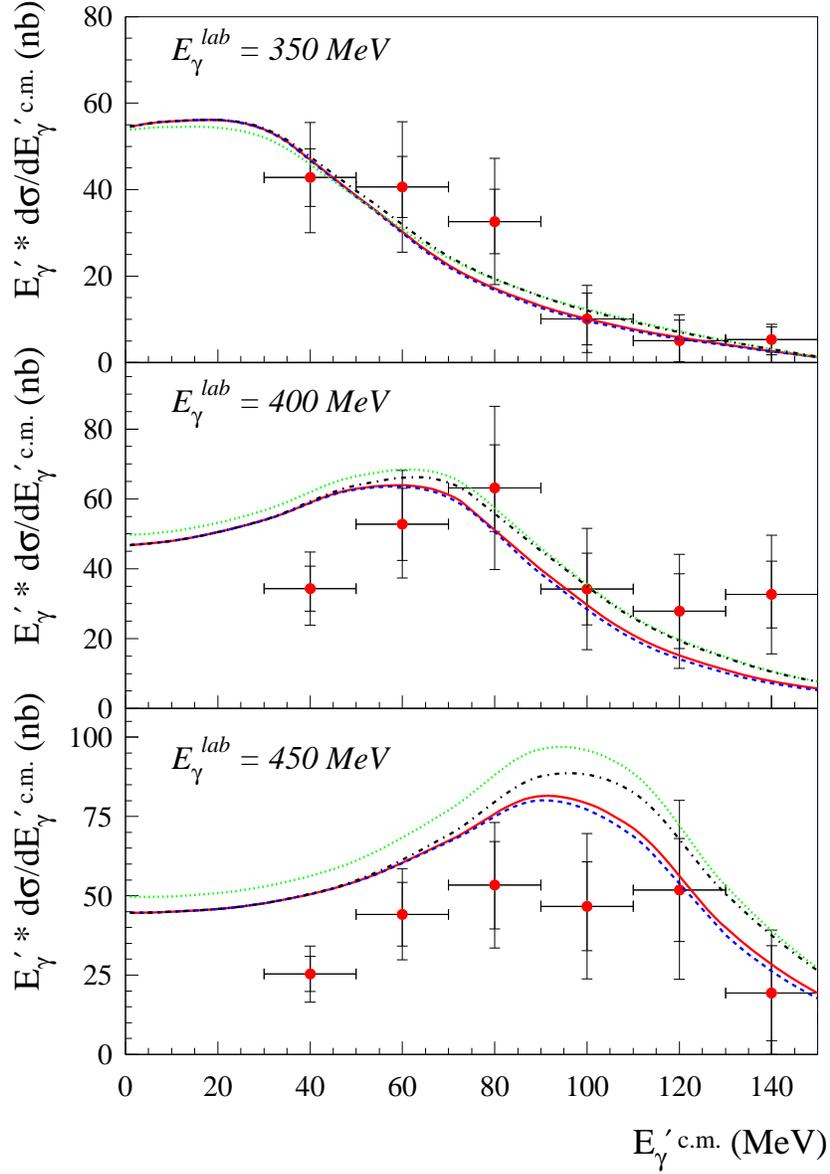}
\caption{
(Color online) Outgoing photon energy dependence of the 
$\gamma p \to \pi^0 p \gamma^\prime$ cross section fully 
integrated over outgoing photon and pion angles for three values of the
incoming photon energy. 
The results of the present  EFT calculation 
are shown for three values of 
$ \mu_{\Delta^+}$~ , in units of $\Delta$ magnetons: 
$ \mu_{\Delta^+} = 1$ (blue dashed curves),  
$\mu_{\Delta^+} = 3$ (red solid curves),  
$ \mu_{\Delta^+} = 5$ (black dashed-dotted curves). 
The green dotted curves show the resonant 
EFT calculation of Ref.~\cite{PV05} for 
$ \mu_{\Delta^+} = 1$. 
The data are from Ref.~\cite{Kotulla:2002cg}. 
}
\figlab{gapio_tot}
\end{figure}

\begin{figure}[t,h]
\includegraphics[width=0.7\columnwidth]{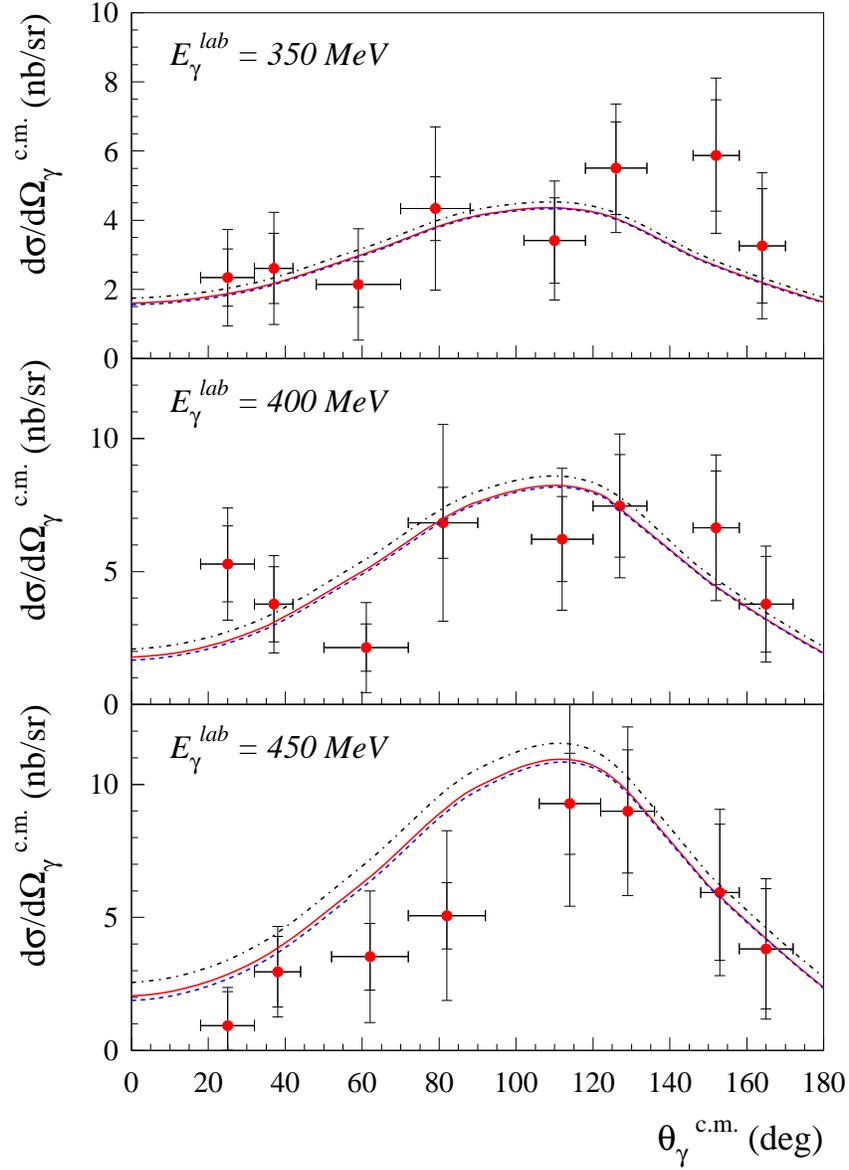}
\caption{
Outgoing photon angular dependence of the 
$\gamma p \to \pi^0 p \gamma^\prime$ {\it c.m.} 
cross section  $d \sigma / d \Omega_\gamma$ 
integrated over outgoing photon energy (for $E_\gamma^\prime \geq 30$~MeV) 
and pion angles for three values of the incoming photon energy. 
The results of the present NLO EFT calculation are shown for three values of 
$\mathrm{Re}\, \mu_{\Delta^+}$ (in units of $\Delta$ magnetons)~: 
$\mathrm{Re}\, \mu_{\Delta^+} = 1$ (blue dashed curves),  
$\mathrm{Re}\, \mu_{\Delta^+} = 3$ (red solid curves),  
$\mathrm{Re}\, \mu_{\Delta^+} = 5$ (black dashed-dotted curves). 
The data are from Ref.~\cite{Kotulla:2002cg}. 
}
\figlab{eftang}
\end{figure}

\begin{figure}[t,h]
\includegraphics[width=0.9\columnwidth]{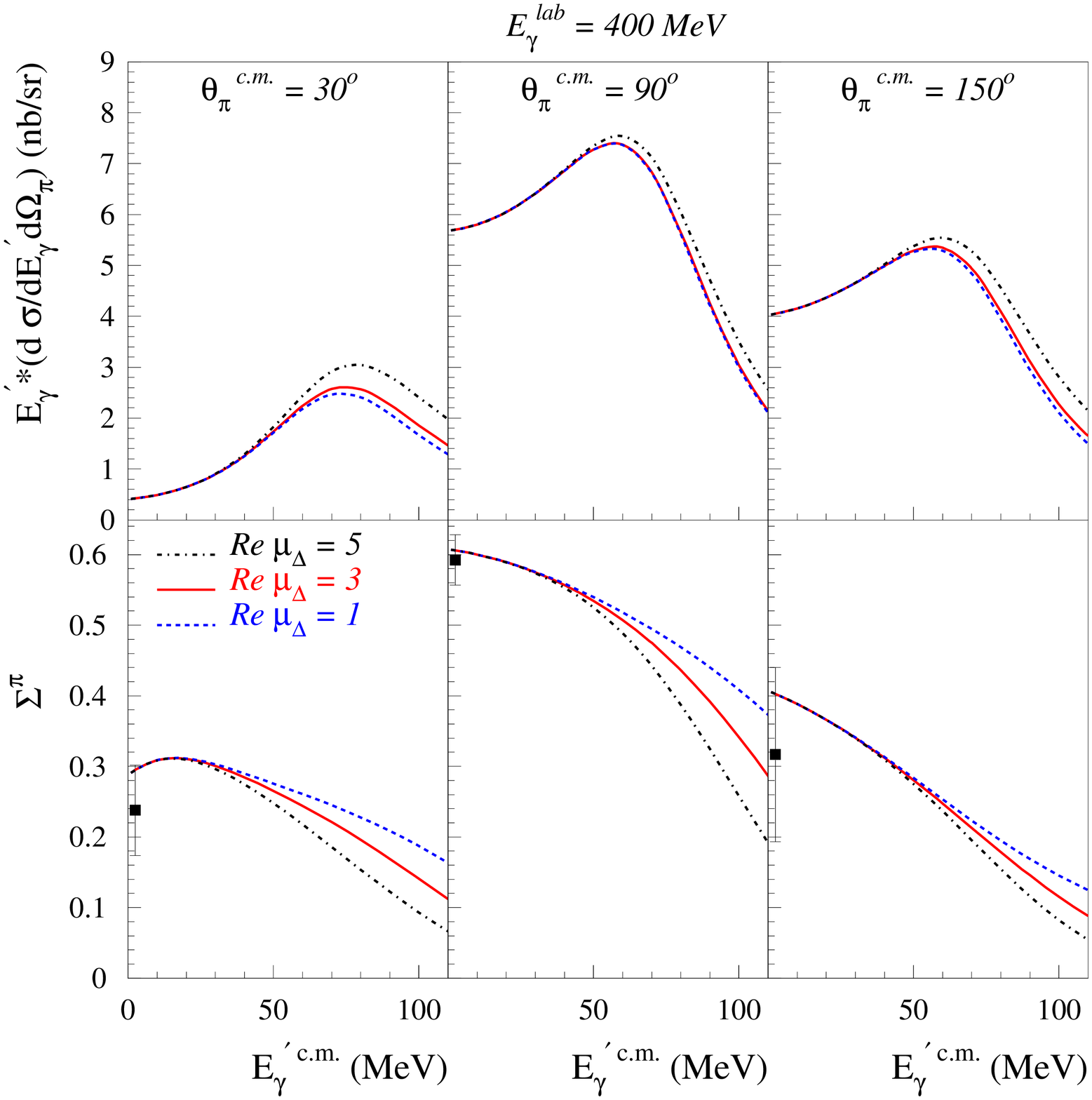}
\caption{
(Color online) Outgoing photon energy dependence of the 
$\gamma p \to \pi^0 p \gamma^\prime$ cross section (top panels) and 
linear photon asymmetry (bottom panels) differential 
with respect to the outgoing photon energy and the pion solid angle 
for three values of the pion polar angle. 
The results of the present NLO EFT calculation are shown for three values of 
$\mathrm{Re}\, \mu_{\Delta^+}$ (in units of $\Delta$ magnetons) 
as indicated on the figure. 
The data points for $E_\gamma^\prime = 0$ show the photon asymmetry data 
of Refs.~\cite{Beck:1999ge,Leukel01} for the $\gamma p \to \pi^0 p$ reaction. 
}
\figlab{eftpi}
\end{figure}

\begin{figure}[t,h]
\includegraphics[width=0.7\columnwidth]{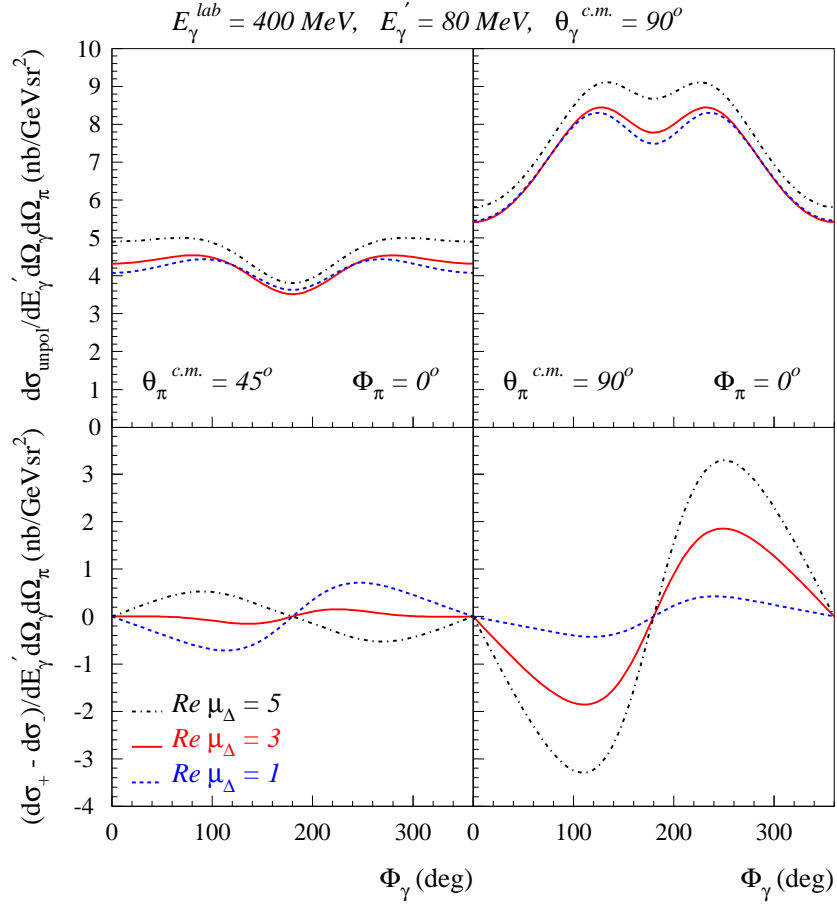}
\caption{
(Color online) Photon azimuthal angular dependence (for $\Phi_\pi = 0^\circ$) 
of the $\gamma p \to \pi^0 p \gamma^\prime$ 5-fold  
differential cross section for an unpolarized photon beam according to 
Eq.~(\ref{eq:unpolexp}) (top panels), and 
of the difference in cross sections for a circularly polarized photon beam  
according to Eq.~(\ref{eq:polexp}) (lower panels), 
for two values of the pion polar angle as indicated on the figure. 
The results of the NLO EFT calculation are shown for three values of 
$\mathrm{Re}\, \mu_{\Delta^+}$~ (in units of $\Delta$ magnetons) 
as indicated on the figure. 
}
\figlab{phieft}
\end{figure}

\begin{figure}[t,h]
\includegraphics[width=0.7\columnwidth]{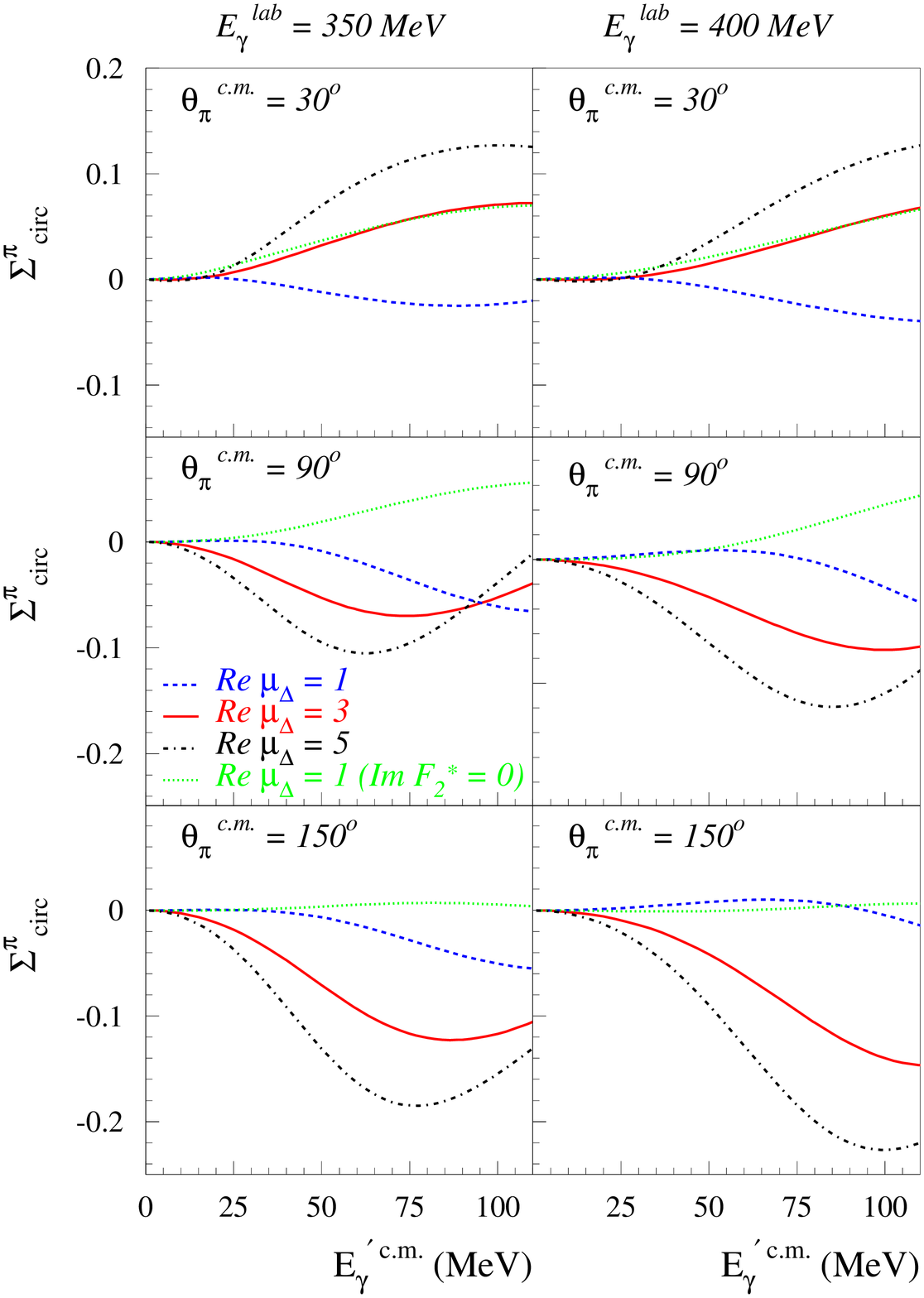}
\caption{
(Color online) Outgoing photon energy dependence of the 
$\gamma p \to \pi^0 p \gamma^\prime$ circular photon asymmetry 
differential with respect to the outgoing photon energy 
and the pion solid angle 
for three values of the pion polar angle and two incoming photon energies as
indicated on the figure. 
The results of the NLO EFT calculation are shown for three values of 
$\mathrm{Re}\, \mu_{\Delta^+}$~ (in units of $\Delta$ magnetons) 
as indicated on the figure. 
The result for $\mathrm{Re}\, \mu_{\Delta^+} = 1$ is  
also shown when turning off the imaginary part of $F_2^\ast$ 
arising from the $\gamma \Delta \Delta$ vertex corrections, 
i.e. for $\mathrm{Im} \, F_2^\ast(0) = 0$ (dotted curves).  
}
\figlab{helpi}
\end{figure}

\begin{figure}[t,h]
\includegraphics[width=0.7\columnwidth]{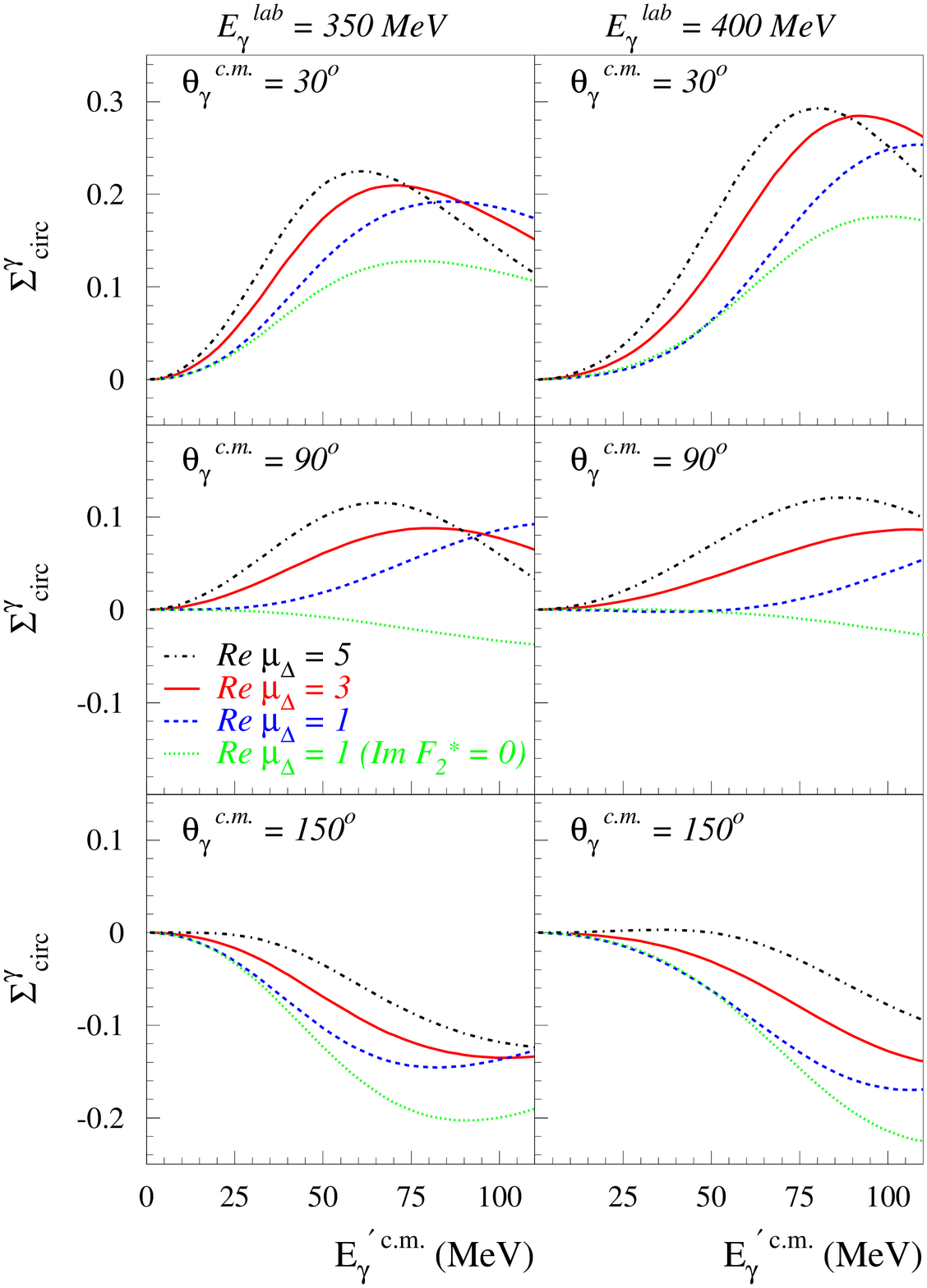}
\caption{
(Color online) Outgoing photon energy dependence of the 
$\gamma p \to \pi^0 p \gamma^\prime$ circular photon asymmetry 
differential with respect to the outgoing photon energy 
and the photon solid angle 
for three values of the photon polar angle and two incoming photon energies as
indicated on the figure. 
The results of the NLO EFT calculation are shown for three values of 
$\mathrm{Re}\, \mu_{\Delta^+}$~ (in units of $\Delta$ magnetons) 
as indicated on the figure. 
The result for $\mathrm{Re}\, \mu_{\Delta^+} = 1$ is  
also shown when turning off the imaginary part of $F_2^\ast$   
arising from the $\gamma \Delta \Delta$ vertex corrections, 
i.e. for $\mathrm{Im} \, F_2^\ast(0) = 0$ (dotted curves).  
}
\figlab{helga}
\end{figure}

\begin{figure}[t,h]
\includegraphics[width=0.7\columnwidth]{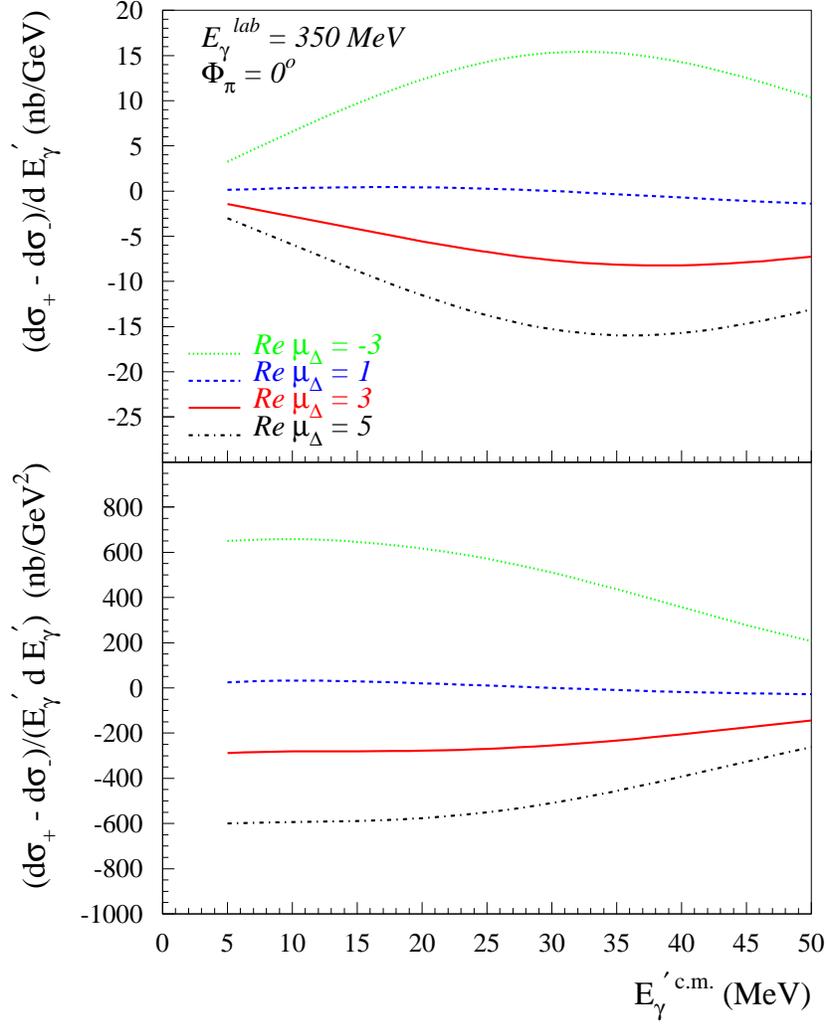}
\caption{
(Color online) Outgoing photon energy dependence of the 
$\gamma p \to \pi^0 p \gamma^\prime$ photon helicity difference cross 
sections,  
differential with respect to the outgoing photon energy, and  
integrated over the full pion polar angular range 
(for $\Phi_\pi = 0^\circ$),   
and over the upper hemisphere for the photon angles. 
The results of the NLO EFT calculation are shown for four values of 
$\mathrm{Re}\, \mu_{\Delta^+}$~ (in units of $\Delta$ magnetons) 
as indicated on the figure. 
}
\figlab{scirc}
\end{figure}

\begin{figure}[t,h]
\includegraphics[width=0.9\columnwidth]{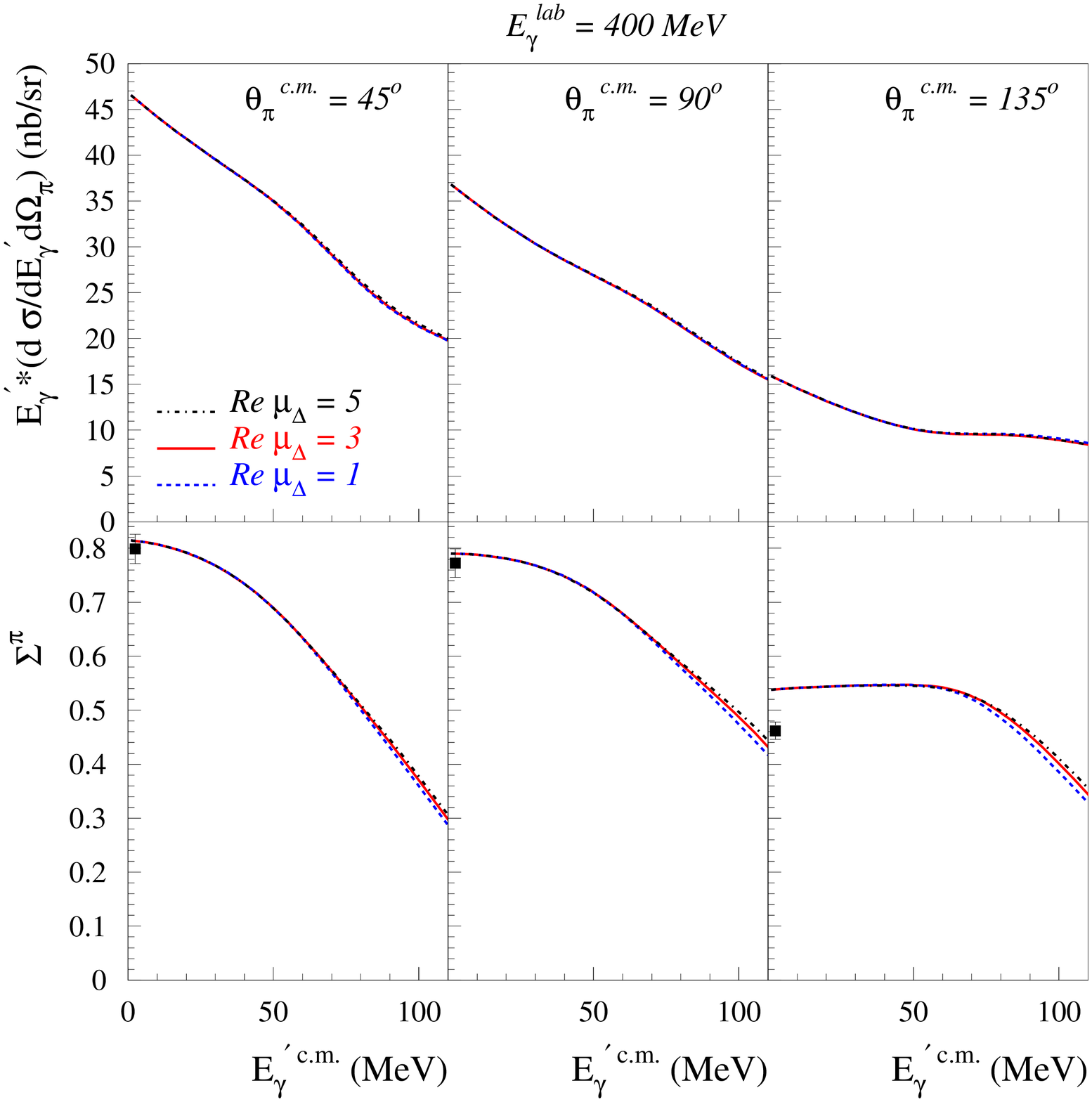}
\caption{
(Color online) Outgoing photon energy dependence of the 
$\gamma p \to \pi^+ n \gamma^\prime$ cross section (top panels) and 
linear photon asymmetry (bottom panels) differential 
with respect to the outgoing photon energy and the pion solid angle 
for three values of the pion polar angle. 
The results of the present NLO EFT calculation are shown for three values of 
$\mathrm{Re}\, \mu_{\Delta^+}$ (in units of $\Delta$ magnetons) 
as indicated on the figure. 
The data points for $E_\gamma^\prime = 0$ show the photon asymmetry data 
of Ref.~\cite{Beck:1999ge} for the $\gamma p \to \pi^+ n$ reaction. 
}
\figlab{eftpip}
\end{figure}

\begin{figure}[t,h]
\includegraphics[width=0.7\columnwidth]{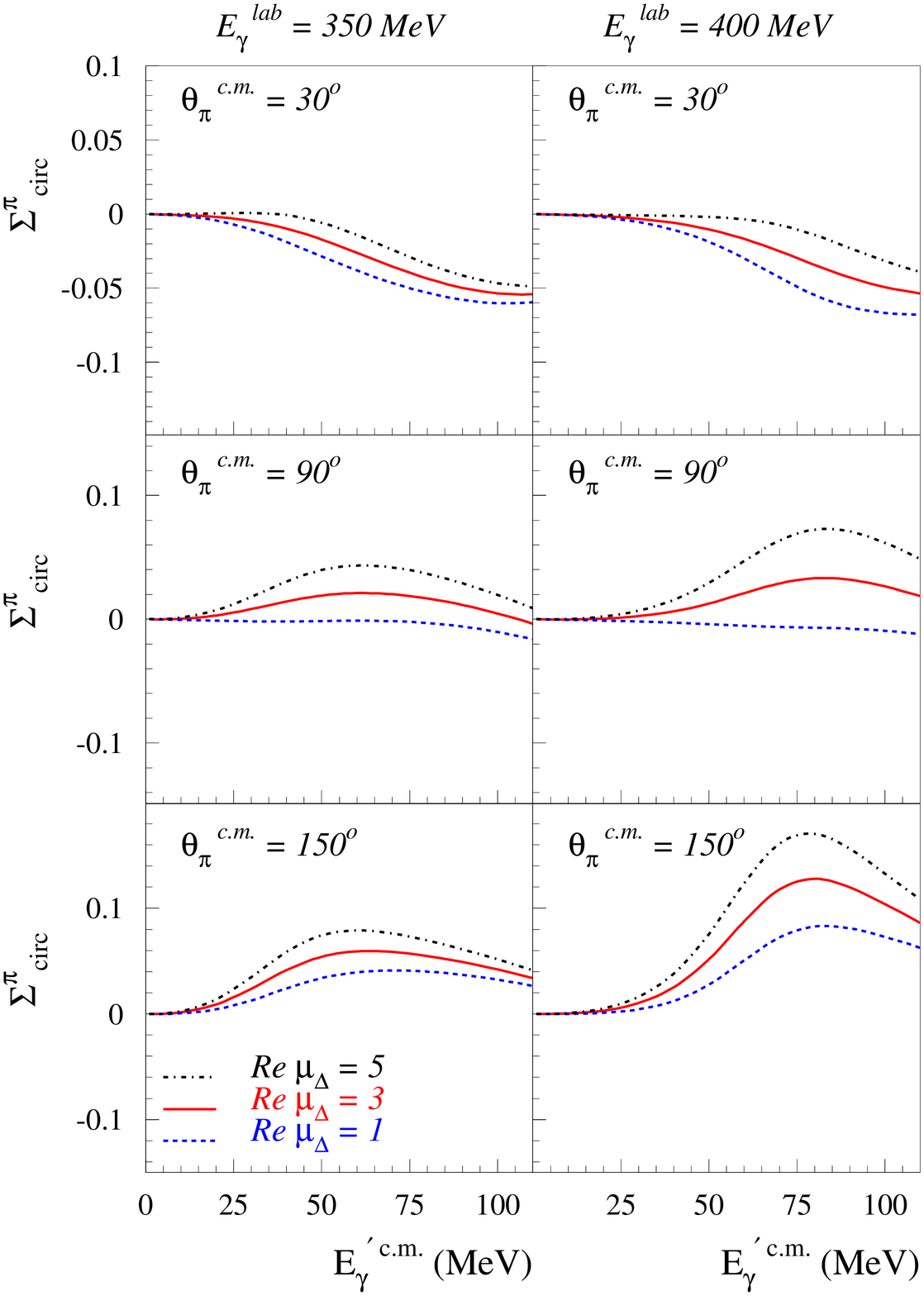}
\caption{
(Color online) Outgoing photon energy dependence of the 
$\gamma p \to \pi^+ n \gamma^\prime$ circular photon asymmetry 
differential with respect to the outgoing photon energy 
and the pion solid angle 
for three values of the pion polar angle and two incoming photon energies as
indicated on the figure. 
The results of the NLO EFT calculation are shown for three values of 
$\mathrm{Re}\, \mu_{\Delta^+}$~ (in units of $\Delta$ magnetons) 
as indicated on the figure. 
}
\figlab{helpip}
\end{figure}

\end{document}